\documentclass[usenatbib,twocolumn]{mnras}
\usepackage[utf8]{inputenc}
\usepackage[T1]{fontenc}
\usepackage{graphicx}
\usepackage{longtable}
\usepackage{wrapfig}
\usepackage{rotating}
\usepackage[normalem]{ulem}
\usepackage{amsmath}
\usepackage{amssymb}
\usepackage{capt-of}
\usepackage{hyperref}
\usepackage{tensor}
\usepackage{amsmath}
\usepackage{caption}
\usepackage{tabularx}
\usepackage{subcaption}
\usepackage{pdfpages}
\usepackage{float}
\usepackage{booktabs}
\usepackage{enumitem}
\usepackage{graphicx}
\usepackage{tensor}
\usepackage{wasysym}
\usepackage{mathtools}
\usepackage{xcolor}
\usepackage{cancel}

\renewcommand{\v}[1]{\boldsymbol{ #1 }}

\DeclarePairedDelimiter{\abs}{|}{|}

\DeclarePairedDelimiter{\p}{(}{)}

\title[Apsidal Angle in MMR]{Apsidal Alignment and Anti-Alignment of
  Planets in Mean-Motion Resonance: Disk-Driven Migration and
  Eccentricity Driving} \author[Laune et al.]{ JT
  Laune,$^{1}$ Laetitia Rodet,$^{1}$ and Dong Lai$^{1}$
  \\
  $^{1}$Department of Astronomy, Cornell Center for Astrophysics and
  Planetary Science, Cornell University, Ithaca, NY 14853, USA \\}
\begin{document}

\maketitle
\begin{abstract}
  Planets migrating in their natal discs can be captured into
  mean-motion resonance (MMR), in which the planets' periods are
  related by integer ratios.  Recent observations indicate that
  planets in MMR can be either apsidally aligned or anti-aligned. How
  these different configurations arise is unclear.  In this paper, we
  study the MMR capture process of migrating planets, focusing on the
  property of the apsidal angles of the captured planets.  We show
  that the standard picture of MMR capture, in which the planets
  undergo convergent migration and experience eccentricity damping due
  to planet-disc interactions, always leads to apsidal anti-alignment
  of the captured planets.  However, when the planets experience
  eccentricity driving from the disc, apsidally aligned configuration
  in MMR can be produced.  In this configuration, both planets’
  resonance angles circulate, but a ``mixed'' resonance angle librates
  and traps the planets near the nominal resonance location. The MMR
  capture process in the presence of disc eccentricity driving is
  generally complex and irregular, and can lead to various outcomes,
  including apsidal alignment and anti-alignment, as well as the
  disruption of the resonance.  We suggest that the two resonant planets
  in the K2-19 system, with their moderate eccentricities and aligned
  apsides, have experienced eccentricity driving from their natal disc
  in the past.
\end{abstract}

\section{Introduction}
\label{sec:org493ee54}
Even before the first detection of exoplanets, it was recognized that
planets can migrate from their initial birth positions due to
interactions with their natal protoplanetary disks \citep[PPDs;][see
\citealt{nelson_planetary_2018} for
review]{lin79_tidal_torques_accret_discs_binar,goldreich_excitation_1979,goldreich_disk-satellite_1980-1}.
The speed and direction of migration depends on the disc density and
temperature profiles.  While planet-disc interactions typically
circularize the planet's orbit, eccentricity excitation can also
occur under some circumstances.
\citet{goldreich03_eccen_evolut_planet_gaseous_disks} demonstrated how
Lindblad and corotation resonances compete to either drive or damp a
giant planet's eccentricity.  More recently,
\citet{teyssandier17_secul_evolut_eccen_protop_discs} and
\citet{ragusa17_eccen_evolut_durin_planet_disc_inter} found that, in
long-term ($\geq 10^4$ orbits) hydrodynamic simulations,
the eccentricity of a large, gap-opening planet exhibits
growth and decay as the system evolves.
\citet{romero21_eccen_drivin_pebbl_accret_low_mass_planet} found that
luminous, super-earth mass protoplanets can be driven to
eccentricities beyond the disk's aspect ratio through a thermal
back-reaction effect from the perturbed gas disk.

As two planets undergo differential migrations in the PPD, they may
encounter mean-motion resonance (MMR), where the planets' periods are
related by integer period ratios $j:j+k$.  For sufficiently slow,
convergent migration, the mutual gravitational interaction between
the planets can lead to MMR capture.  The period ratio distribution of
super-Earths/mini-Neptunes discovered by the Kepler mission indeed
shows an excess of planet pairs near MMRs, although the observed MMR
occurrence rate is much lower than the MMR capture rate predicted
using the simplest migration/capture model.  The ``near MMR'' systems
typically have period ratios slightly larger (by 1-2\%) than exact
resonance \citep[e.g.][]{fabrycky_architecture_2014}.  These
discrepancies could be explained by the instability of the captured
state during disk-driven migration
\citep{goldreich_overstable_2014,deck_migration_2015,delisle15_stabil_reson_config_durin_migrat,xu16_disrup_planet_orbit_throug_evect,xu_migration_2018},
resonance repulsion due to tidal eccentricity damping or planet-disk
interactions
\citep{lithwick12_reson_repul_planet_pairs,batygin_analytical_2013,delisle14_tidal_dissip_format_of_reson_planet,choksi_sub-neptune_2020},
late-time dynamical instability \citep{izidoro_breaking_2017}, and/or
outward (divergent) migration due to planetesimal scatterings
\citep{chatterjee15_planet_inter_can_explain_myster}.

An interesting property of MMR capture concerns the relative
apsidal angle $\Delta\varpi=\varpi_1-\varpi_2$ of the captured planets.
For the first-order MMR ($j:j+1$), the two resonant angles are
\begin{align}
\label{circangles1}
 \theta_1 &= (j+1)\lambda_2 - j\lambda_1 - \varpi_1, \\
\label{circangles2}
 \theta_2 &= (j+1)\lambda_2 - j\lambda_1 - \varpi_2,
\end{align}
where $\lambda_1$ and $\lambda_2$ are the mean longitudes of the planets.
In equilibrium, we expect $\theta_1$ and $\theta_2$ to be either
$0^\circ$ or $180^\circ$, implying $\Delta\varpi=0^\circ$ or
$180^\circ$. As we will see in Section~\ref{sec:org985dec7}, the
conventional treatment of migration and resonance capture in PPDs always
produces apsidal anti-alignment when the planet's eccentricity is damped by the disk.

Only a handful of systems in or near resonance have constrained
measurements of $\Delta\varpi$, and, indeed, most exhibit apsidal
anti-alignment ($\varpi=180^\circ$).  Kepler-88 hosts two planets, b
and c, near the 1:2 resonance with periods 10.9 d and 22.3 d and
masses $214M_\oplus$ and $9.5M_\oplus$, respectively
\citep{weiss_discovery_2020}.  Kepler-9b and c are two sub-Jupiter
mass planets ($M_b=0.25M_J$ and $M_c=0.17M_J$) orbiting near the 1:2
MMR \citep[$P_b=19.2$ d and $P_c=38.9$ d;][]{holman10_kepler}.  Both of
these systems are constrained by their photodynamical data to have
$\Delta\varpi\approx180^\circ$ \citep{antoniadou_exploiting_2020}.
The system K2-24 has two approximately Neptune-mass planets
($M_b=19M_\oplus$, $M_c=15.4M_\oplus$), K2-24b and c, near the 1:2
resonance with periods $P_b=20.8$ d and $P_c = 42.3$ d
\citep{petigura18_dynam_format_near_reson_k2_system}.  The
observations of K2-24 are consistent with either the apsidally aligned
or anti-aligned configuration \citep{antoniadou_exploiting_2020}.

The only aligned system detected thus far is K2-19, a three planet
system around a K dwarf star
\citep[\(M=0.88M_\odot\);][]{armstrong15_one_closes_exopl_pairs_to}.
The planets K2-19b and c are near the 2:3 period ratio (\(P_b=7.9\) d,
\(P_c=11.9\) d), and planet K2-19d lies on an orbit interior to the
other two at \(P_d=2.5\) d. Their masses are $M_d<10 M_\oplus$,
$M_b=32.4 M_\oplus$, and $M_c=10.8 M_\oplus$, with radii
$R_d\approx 1.1 R_\oplus$, $R_b\approx 7.0R_\oplus$, and
$R_c\approx 4.1R_\oplus$.  The K2-19 photometry data indicates that the
innermost planet has $e_d\simeq 0$, planets b and c
have moderate eccentricities, and
\(e_b\approx e_c \approx 0.2\). Their apsidal angles are constrained to
within a few degrees of 0$^\circ$
\citep{petigura_k2-19b_2020,petit_resonance_2020}.
The origin of this alignment is unclear.

Investigating how the two planets in K2-19 could have formed with
\(\Delta\varpi=0^\circ\) through resonance capture and mutual
migration (while other systems have $\Delta\varpi=180^\circ$) may
offer us new insight into its dynamical history as well as a better
understanding of the genesis of extrasolar orbital configurations in
general.  In this paper, we study the apsidal property of planets in
MMRs produced by disk-driven migration.  In Section
\ref{sec:org985dec7}, we present the standard picture of resonant
capture and explore the parameter space for the coupling between the
planets and the protoplanetary disk.  We show that as long as the disk
damps the planetary eccentricity, MMR capture always leads to apsidal
anti-alignment.  In Section~\ref{sec:org4c72d92}, we explore the
apsidal property for a test particle in the vicinity of the MMR with
an eccentric planet's MMR. These results guide our analysis in Section
\ref{sec:orgd5c121f}, where we show that when planet-disc interaction
drives the planet's eccentricity toward a finite, non-zero value,
apsidal alignment can be produced under certain circumstances. We
conclude in Section~\ref{sec:org71c2a7a}.

\section{Disk-Driven MMR Capture: Standard Apsidal Architecture}
\label{sec:org985dec7}
Consider two planets with masses $m_1$ and $m_2$ orbiting
a star of mass $M$ (set to 1$M_\odot$ throughout this paper)
with semimajor axes $a_1$ and $a_2$ ($>a_1$) in a gaseous disk.
At low eccentricities, the orbital decay rate and eccentricity
damping rate due to planet-disc interactions are given by
\begin{align}
  \label{eq:disforce1}
  \frac{\dot{a}_i}{a_i} &= -\frac{1}{T_{m,i}} -\frac{2e_i^2}{T_{e,i}},\\
  \label{eq:disforce}
  \frac{\dot{e}_i}{e_i} &= -\frac{1}{T_{e,i}},
\end{align}
\noindent
for each planet ($i=1,2$). For low-mass planets undergoing type-I
migration and typical disc profiles, we have
\citep{tanaka_three-dimensional_2004,cresswell_three-dimensional_2008}
\begin{align}
  \label{eq:TePhys}
  T_{e,i} &\sim \frac{M^2}{\Sigma a_i^2 m} h^4 n_i^{-1}, \\
  \label{eq:TmPhys}
  |T_{m,i}| &\simeq \frac{T_e}{3.46 h^2} ,
\end{align}
\noindent
where $a_i$ is the semimajor axis (SMA), $\Sigma$ is the disc surface
density, $h$ is the aspect ratio of the disk, and $n_i$ is the mean
motion. Note that $T_{e,1}/T_{e,2}=m_2/m_1=1/q$.
Throughout the paper, we treat $T_{e,1}$ and $T_{e,2}$ as constants
(independent of the semimajor axis) and use
$T_{e,0} = \sqrt{T_{e,1}T_{e,2}}$ as the basic timescale for
eccentricity damping. We set $h=0.03$ and $T_{e,0}=1000P_{2,0}$, where
$P_{2,0}$ is the initial period of $m_2$.

\subsection{Equations of motion}
\label{sec:orgf761dbc}
\begin{figure*}
  \centering
  \includegraphics[width=0.7\textwidth]{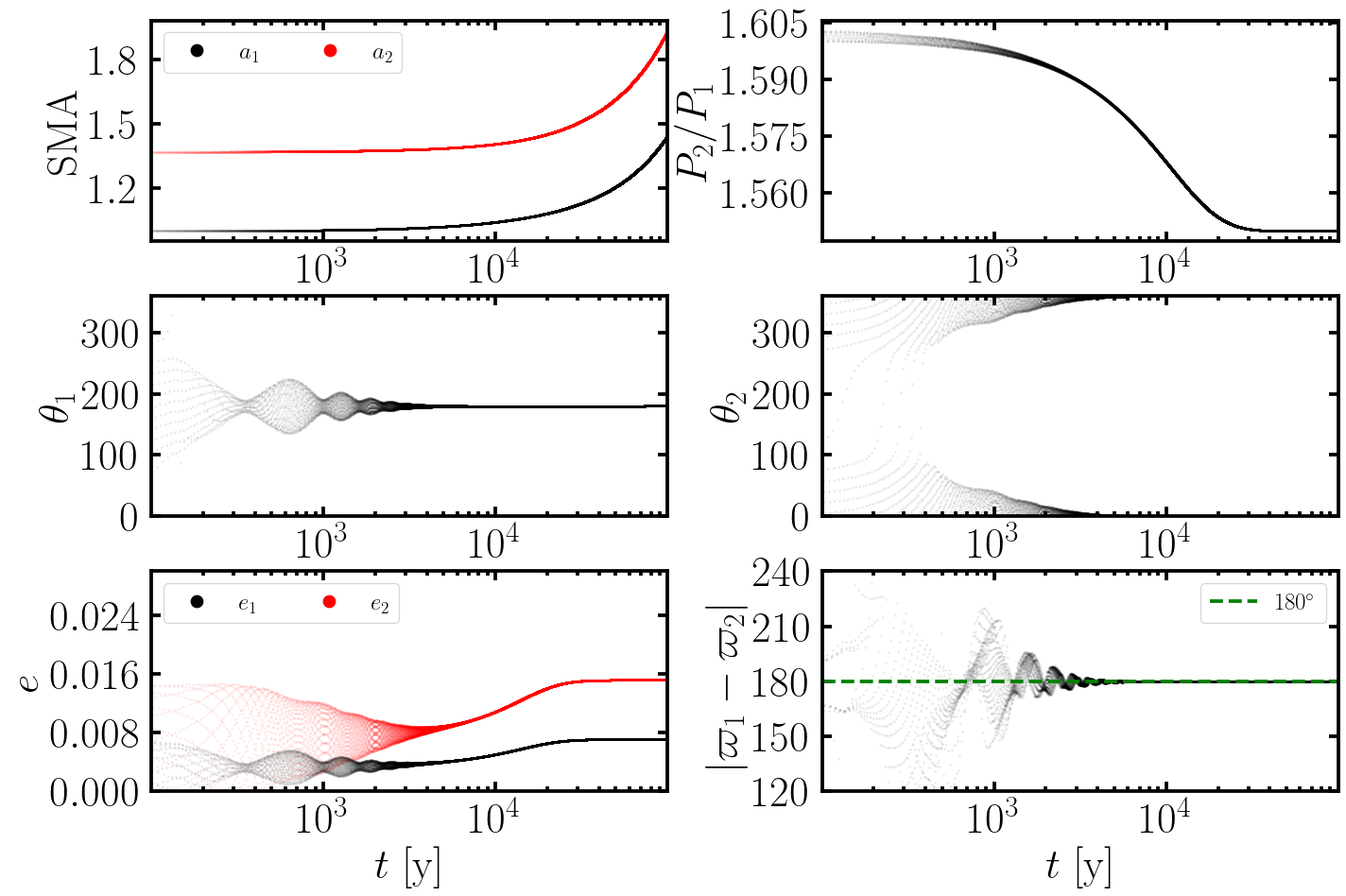}
  \caption{Standard MMR capture process for $h=0.03$ and
    $q=m_1/m_2=2$. The inner planet starts at the SMA $a_1=1$ au and
    the outer planet starts wide of the 2:3 resonance at
    $P_2/P_1=1.6$.  Both planets have very small initial
    eccentricities, $e_1=e_2=0.001$.  Note that to have convergent
    migration, we adopt $T_{m,i}<0$ (i.e.  both planets migrate
    outwards).  The planets are captured into the resonance around
    $t=20,000$ yrs, indicated by the libration of
    $\theta_1\to180^\circ$ and $\theta_2\to 0^\circ$ and the period
    ratio approaching the equilibrium value ($\simeq 1.55$).  While in
    resonance, the eccentricities are driven to the equilibrium
    values, $e_1\approx 0.008$ and $e_2\approx 0.016$, and the
    periapses are anti-aligned.}
  \label{fig:standardex}
\end{figure*}

For two planets on coplanar orbits near the $j:j+1$ MMR, the
Hamiltonian can be approximated to order \(\mathcal{O}(e^2)\) as
\citep{murray_solar_2000}
\begin{align}
  \label{hamiltonian}
  H=H_{\rm kep}+H_{\rm res}+H_{\rm sec},
\end{align}
with the Keplerian, resonant, and secular terms given by
\begin{align}
  H_{\rm kep} = & -\frac{G M m_{1}}{2 a_{1}}-\frac{G M m_{2}}{2 a_{2}},\\
  \label{eq:Hcompmassres}
  H_{\rm res} = & -\frac{G m_{1} m_{2}}{a_{2}}
                  \left[
                  f_{1} e_{1} \cos \theta_{1} 
                  +f_{2} e_{2} \cos \theta_{2}\right],\\
  H_{\rm sec} = &-\frac{G m_{1} m_{2}}{a_{2}}\left[f_{3} (e_1^2 + e_2^2)
                  +f_4e_1e_2\cos(\varpi_2-\varpi_1)
                  \right], 
\end{align}
\noindent
where \(\theta_1\) and \(\theta_2\) are given in equations
\eqref{circangles1} and \eqref{circangles2}.
Here, the \(f_i\) are functions of the semimajor
axis ratio \(\alpha=a_1/a_2\) that can be found in Appendix B of
\citet{murray_solar_2000}:
\begin{align}
\label{eq:coefficients1}
  f_1 &= \frac12[2(j+1)+\alpha D]b_{1/2}^{(j+1)}(\alpha), \\
\label{eq:coefficients2}
  f_2 &= -\frac12[2j+1+\alpha D]b_{1/2}^{(j)}(\alpha), \\
  f_3 &= \frac18[2\alpha D + \alpha^2 D^2]b_{1/2}^{(0)}(\alpha), \\
  f_4 &= \frac14[2-2\alpha D - \alpha^2 D^2]b_{1/2}^{(1)}(\alpha),
\end{align}
\noindent where \(D\equiv d/d\alpha\)
and the \(b_{l}^m\) are Laplace coefficients.
Near the 2:3 MMR, we have $f_1 \approx 2.0$, $f_2\approx -2.5$,
$f_3\approx 1.15$ and $f_4\approx 2.0$.

The Hamiltonian system defined by equation~\eqref{hamiltonian} admits
eight coupled ordinary differential equations, which we may integrate
together with dissipative terms (equations~\ref{eq:disforce1} and
\ref{eq:disforce}) to simulate MMR capture.  The canonical Poincair\'e
momentum-coordinate variables are
\begin{align}
  \Lambda_i = m_i\sqrt{GMa_i}, &\quad \lambda_i;\\
  \Gamma_i = \Lambda_i(1-\sqrt{1-e_i^2})\simeq \frac12\Lambda_ie_i^2, &\quad -\varpi_i.
\end{align}
\noindent We apply Hamilton's equations to generate the equations of
motion (to second order in eccentricities) and add in the dissipative
effects:
\begin{align}
  \label{eq:a1dot}
  \frac{\dot a_1}{a_1} &= \frac{2j  n_2}{\sqrt\alpha}(f_1e_1\sin\theta_1+f_2e_2\sin\theta_2)
                         -\frac{1}{T_{m,1}}-\frac{e_1^2}{T_{e,1}},
\\
  \label{eq:a2dot}
  \frac{\dot a_2}{a_2} &= -2(j+1)\mu_1 n_2(f_1e_1\sin\theta_1+f_2e_2\sin\theta_2)
                         -\frac{1}{T_{m,2}}-\frac{e_2^2}{T_{e,2}},
\\
  \label{eq:e1dot}
  \dot e_1 &= -\frac{\mu_2n_2 f_1}{\sqrt{\alpha}}\sin\theta_1 
             + \frac{\mu_2f_4e_2n_2}{\sqrt\alpha}\sin(\varpi_1-\varpi_2)
             - \frac{e_1}{T_{e,1}}
             ,
\\
  \label{eq:e2dot}
  \dot e_2 &= - \mu_1 n_2 f_2 \sin\theta_2 
             - \mu_1 f_4 e_1 n_2\sin(\varpi_1-\varpi_2)
             - \frac{e_2}{T_{e,2}}
             ,
\\
  \label{eq:l1dot}
  \dot\lambda_1 &= n_1 + \frac{\mu_2n_2}{2\sqrt\alpha}f_1 e_1 \cos\theta_1 \nonumber\\
  &+\frac{\mu_2 n_2}{\sqrt\alpha}\left(f_3 e_1^2 + \frac{f_4 e_1e_2}{2}\cos(\varpi_2-\varpi_1)\right),
\\
  \label{eq:l2dot}
  \dot\lambda_2 &= n_2 + \mu_1n_2\left(2f_1e_1\cos\theta_1+ \frac52 f_2 e_2\cos\theta_2\right) \nonumber\\
  &+\mu_1n_2\left(2f_3e_1^2 + 3f_3e_2^2 + \frac52 f_4e_1e_2\cos(\varpi_2-\varpi_1)\right),
\\
  \label{eq:pom1dot}
  \dot\varpi_1 &= \frac{\mu_2n_2}{\sqrt\alpha e_1}f_1\cos\theta_1 
  +\frac{\mu_2n_2}{\sqrt\alpha}\left(2f_3+f_4\frac{e_2}{e_1}\right),
\\
  \label{eq:pom2dot}
  \dot\varpi_2 &= \frac{\mu_1 n_2}{e_2}f_2\cos\theta_2
  +\mu_1n_2\left(2f_3+f_4\frac{e_1}{e_2}\right).
\end{align}
\noindent
Note that the right-hand sides of these equations depend on
$\lambda_1$ and $\lambda_2$ solely through the combination
$(j+1)\lambda_2-j\lambda_1$.  By combining equations~\eqref{eq:l1dot}
and \eqref{eq:l2dot} into a single equation for
$(j+1)\dot\lambda_2-j\dot\lambda_1$, we may reduce the number of
equations to seven.  We integrate the system with the Runge-Kutta
method of order 5(4) included in the \(\mathtt{scipy}\)
Python package.  We set a relative and absolute tolerance
\(\epsilon=10^{-9}\).  The resonance angles are initialized over a
uniform distribution between \(0^\circ\) and \(360^\circ\).  At
\(t=0\), we set \(a_1=1~\mathrm{au}\), \(P_2/P_1=1.6\), and
\(e_1=e_2=0.001\).

An example of MMR capture is shown in Fig.~\ref{fig:standardex}.  The
period ratio \(P_2/P_1\) starts wide of the nominal 2:3 resonance
value.  After around \(2~\rm{kyr}\) of convergent migration, the planets
are caught into the MMR, indicated by the stabilization of \(\theta_1\) to
\(180^\circ\) and \(\theta_2\) to \(0^\circ\).  The planets' eccentricities
level off at their equilibrium values at \(e_1\approx 0.008\) and
\(e_2\approx0.016\), and the planets become apsidally anti-aligned with
\(\varpi_1-\varpi_2\approx 180^\circ\).

The final period ratio $P_2/P_1\simeq 1.55$ differs from 3/2, which
can be explained by the equations for
$\dot\theta_1=(j+1)\dot\lambda_2-j\dot\lambda_1-\dot\varpi_1$ and
$\dot\theta_2=(j+1)\dot\lambda_2-j\dot\lambda_1-\dot\varpi_2$.  If we
truncate the equations to lowest order in eccentricity
($\mathcal O(e_i^{-1})$), the dynamics are dominated by the
$\dot\varpi_i$ terms (equations~\ref{eq:pom1dot} and
\ref{eq:pom2dot}), and we have
\begin{align}
  \label{eq:theta1dot}
  \dot\theta_1&\simeq(j+1)n_2-jn_1-\frac{\mu_2n_2}{\sqrt\alpha e_1}|f_1|\cos\theta_1,
  \\
  \label{eq:theta1dot}
  \dot\theta_2&\simeq(j+1)n_2-jn_1+\frac{\mu_1 n_2}{e_2}|f_2|\cos\theta_2.
\end{align}
Hence, in equilibrium, it is expected that $P_2/P_1=n_1/n_2>(j+1)/j=1.5$
for the observed values of $\theta_1=180^\circ$ and $\theta_2=0^\circ$.

\subsection{Equilibrium}
\label{sec:org8ceeefb}
\begin{figure}
  \centering
  \begin{subfigure}[t]{0.225\textwidth}
  \includegraphics[width=1\textwidth]{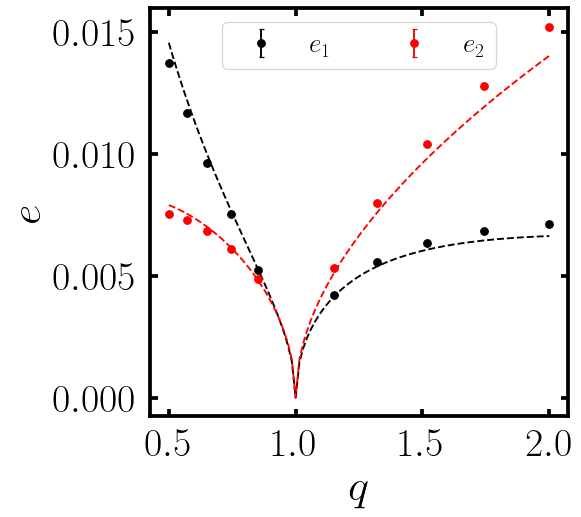}
  \caption{ }
  \label{fig:standardeqecc}
  \end{subfigure}
  \begin{subfigure}[t]{0.225\textwidth}
  \includegraphics[width=1\textwidth]{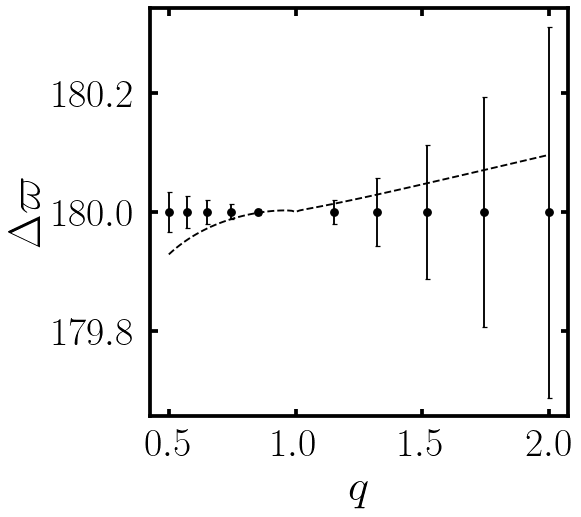}
  \caption{ }
  \label{fig:standardDpom}
  \end{subfigure}
  \caption{Equilibrium eccentricities $e_1, e_2$ (left) and relative
    apsidal angle $\Delta\varpi=\varpi_1-\varpi_2$ (right) for two
    planets captured into the 2:3 MMR, plotted against the mass ratio
    $q=m_1/m_2$.  The dashed lines are analytical results determined
    by solving equations~(\ref{dote1}) -- (\ref{dotdpom}) and
    (\ref{doteta}).  The dots are obtained by integrating the
    time-dependent equations of motion and time-averaging the
    eccentricities over the last 10\% of the integration.  We hold
    $h=0.03$ and $T_{e,0}=\sqrt{T_{e,1}T_{e,2}}=1000$ yrs constant,
    but allow the eccentricity damping and migration timescales to
    vary with $q$, as in equations~\eqref{eq:TePhys} and
    \eqref{eq:TmPhys}.  }
\label{fig:standard}
\end{figure}

The MMR capture shown in Fig.~\ref{fig:standardex} leads to an equilibrium
state in period ratio, resonant angles, eccentricities, and
\(\Delta\varpi\).  By requiring $\dot e_1=0$, $\dot e_2=0$, and
$\dot\varpi_1-\dot\varpi_2=0$, we arrive at three
equations:
\begin{equation}
\label{dote1}
  \dot e_1 = -\frac{\mu_2n_2}{\sqrt\alpha} [f_1\sin\theta_1 - f_4e_2 \sin(\varpi_1-\varpi_2)] - \frac{e_1}{T_{e,1}}=0
\end{equation}

\begin{equation}
\label{dote2}
  \dot e_2 = -q\mu_2n_2 [f_2\sin\theta_2 + f_4e_1 \sin(\varpi_1-\varpi_2)]- \frac{e_2}{T_{e,2}}=0
\end{equation}

\begin{align}
\label{dotdpom}
  \frac{d}{dt}\Delta\varpi = \dot\varpi_1-\dot\varpi_2
  &= \mu_2n_2 \left[ \frac{f_1\cos\theta_1}{\sqrt{\alpha} e_1}
     - \frac{qf_2\cos\theta_2}{e_2}\right.\nonumber \\
  &\quad+ \left.\frac{2f_3}{\sqrt{\alpha}} + \frac{f_4e_2}{\sqrt\alpha e_1}
    - 2qf_3 - \frac{qf_4e_1}{ e_2}\right]=0.
\end{align}

\noindent To first order in eccentricity, the first two
equations determine the equilibrium values of \(\theta_1\) and
\(\theta_2\), while the last implies that \(e_2/e_1 \sim q=m_1/m_2\).
In the absence of dissipation, the following
quantities are strictly conserved \citep[e.g.][]{xu_migration_2018}:
\begin{align}
  J &= \Lambda_1\sqrt{1-e_1^2} + \Lambda_2\sqrt{1-e_2^2},\\
  K &= \frac{j+1}{j} \Lambda_1 + \Lambda_2.
\end{align}
\noindent
Following \citet{xu_migration_2018}, we define 
\begin{align}
  \eta(\alpha, e_1, e_2) &\equiv - 2(q/\alpha_0+1)\p*{\frac{J}{K}-\left.\frac{J}{K}\right|_{0}},
\end{align}
\noindent
where \(\alpha_0 = [j/(j+1)]^{2/3}\) and \(\left.(J/K)\right|_{0}\) is
evaluated at \(e_i=0\) and \(\alpha=\alpha_0\).
For $|\alpha-\alpha_0|\ll 1$ and $e_1,e_2\ll 1$, we have
\begin{align}
  \eta \simeq \frac{q(\alpha-\alpha_0)}{j\sqrt{\alpha_0}(q/\alpha_0+1)} + q\sqrt{\alpha_0}e_1^2 + e_2^2.
\end{align}
\noindent
Since \(\eta\) is conserved in the absence of dissipation, 
the only nonzero terms in its derivative, \(\dot{\eta}\),
can be from the dissipative effects.
In equilibrium, we require $\dot\eta = 0$, i.e.
\begin{align}
\label{doteta}
  \dot\eta = \frac{q\alpha}{j\sqrt{\alpha_0}(q\alpha_0^{-1}+1)}&\left[ \frac{1}{T_{m,2}} - \frac{1}{T_{m,1}}
      + \frac{2e_1^2}{T_{e,1}}- \frac{2e_2^2}{T_{e,2}} \right] \nonumber\\
    &- q\alpha_0^{1/2}\frac{2e_1^2}{T_{e,1}} - \frac{2e_2^2}{T_{e,2}}=0.
\end{align}
\noindent We note that \(\dot\eta\) depends only on the
\emph{effective} migration rate,
\(1/T_m \equiv 1/T_{m,2} - 1/T_{m,1}\).

Utilizing the $\mathtt{scipy}$ root finding library, assuming
$\alpha=\alpha_0$, we solve the four equations~\eqref{dote1} --
\eqref{dotdpom} and \eqref{doteta} for the equilibrium values of
$\theta_1, \theta_2, e_1, e_2$.  The equilibrium \(e_i\)'s and
\(\Delta\varpi\)'s for comparable mass planets \((q\in[0.5,2])\) are
given in Figures \ref{fig:standardeqecc} and \ref{fig:standardDpom} as
the dashed lines.  The equilibrium eccentricities go approximately as
\(e_2/e_1 \sim q\), and all systems are predicted to have
$\Delta\varpi\approx180^\circ$.  To validate these analytical results,
we also integrate the time dependent equations which simulate MMR
capture and plot the average \(e_1\), \(e_2\), and \(\Delta\varpi\)
over the last 10\% of the simulation.  For \(T_{e,1}<T_{e,2}\)
($T_{e,1}>T_{e,2}$), we set \(T_{m,i}<0\) ($T_{m,i}>0$), corresponding
to outward (inward) migration. The numerical (markers) and analytical
(dashed lines) results largely agree.  Thus, in the standard picture
of MMR capture in PPDs, comparable mass planets always end up settling
down to \(\theta_1\approx180^\circ\) and \(\theta_2\approx0^\circ\),
and the eccentricity vectors of the planets are anti-aligned,
$|\Delta\varpi|=|\varpi_1-\varpi_2|=|\theta_2-\theta_1|=180^\circ$.

\subsection{Survey of parameter space: Eccentricity damping timescales}
\label{sec:org5a31b76}
\begin{figure}
  \centering
  \includegraphics[width=0.3\textwidth]{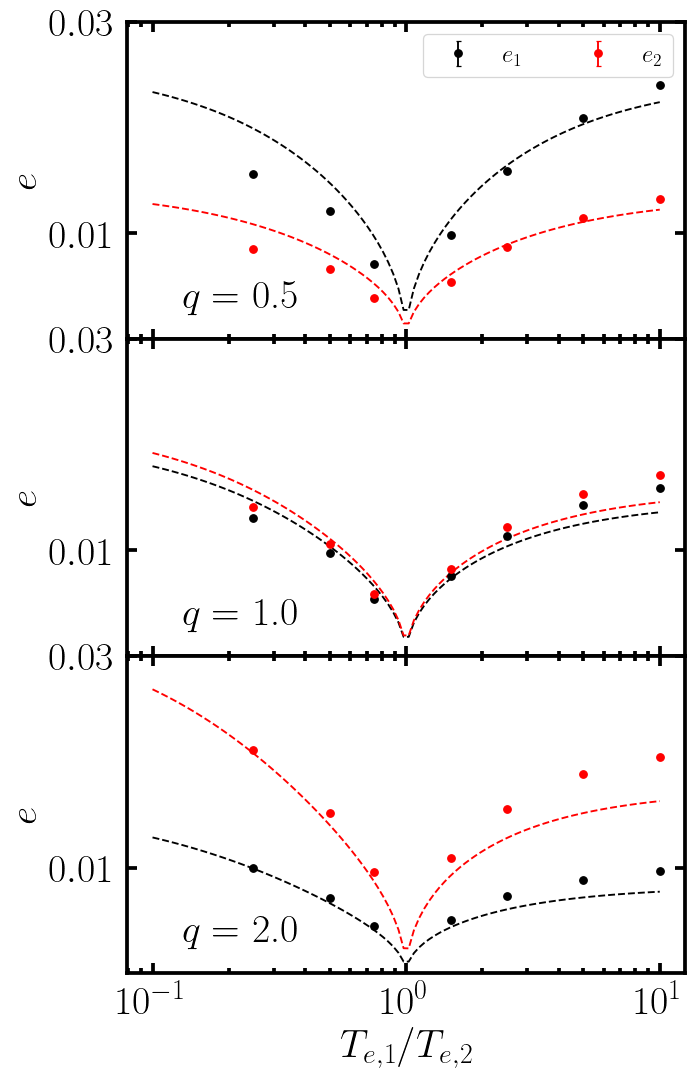}
  \caption{ The final time-averaged eccentricities (dots) of the two
    planets captured in the 2:3 MMR as a function of $T_{e,1}/T_{e,2}$
    for $q=m_1/m_2=0.5, 1,$ and $2$. We have kept $T_{e,0}$, $h$ and the
    initial conditions fixed to the same values as in Figure
    \ref{fig:standardex}.  The dashed lines indicate the analytical
    equilibrium eccentricities obtained by solving equations
    (\ref{dote1}) -- (\ref{dotdpom}) and (\ref{doteta}), as in
    Fig.~\ref{fig:standardeqecc}.  For $T_{e,1}/T_{e,2}>1$
    ($T_{e,1}/T_{e,2}<1$), migration is inward (outward), since we
    keep the ratio $T_{e,i}/|T_{m,i}|\propto h^2$. For $q=0.5$, the
    inward migrating branch ($T_{e,1}>T_{e,2}$) agrees well with the equilibrium result.
    However, for the outward migrating branch the analytic
    solution typically overestimates the final eccentricities. The
    results are similar for $q=1,2$.}
  \label{fig:eqecc}
\end{figure}

\begin{figure}
  \centering
  \includegraphics[width=0.3\textwidth]{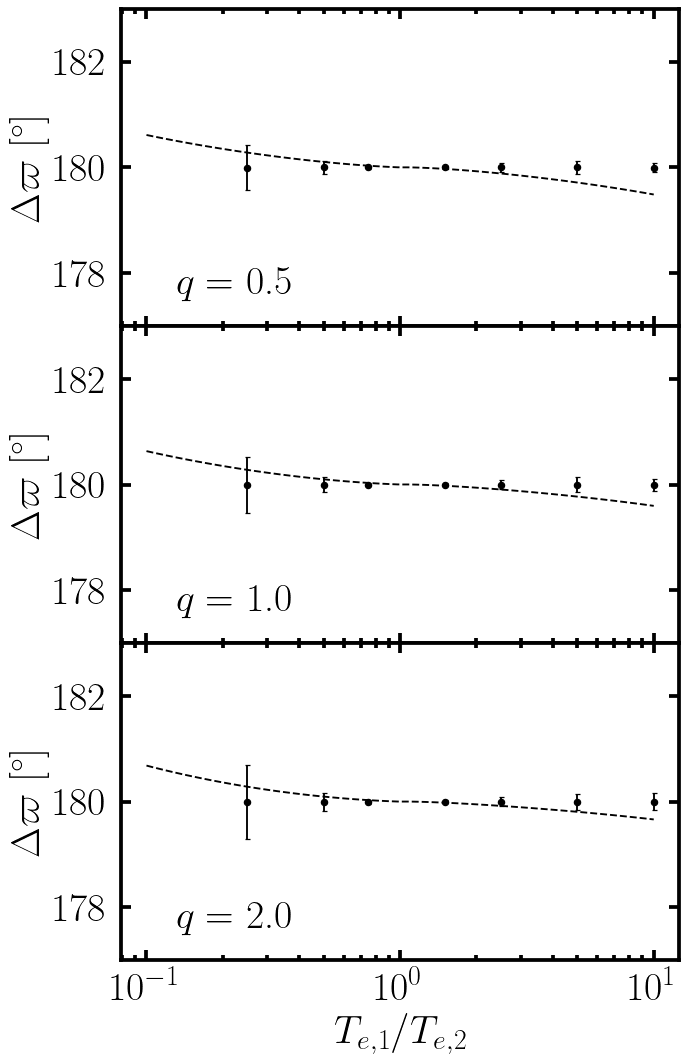}
  \caption{Same as Fig.~\ref{fig:eqecc} but showing $\Delta\varpi$. In
    all cases the apsides of the two planets are anti-aligned.
  }
  \label{fig:eqDpom}
\end{figure}

In Section~\ref{sec:org8ceeefb}, we adopted the standard scaling
relation -- with \(T_{e,1}/T_{e,2} = 1/q=m_2/m_1\) and
\(T_{e,i}=3.46h^2|T_{m,i}|\) -- which always gives rise to apsidal
anti-alignment for typical disc conditions (\(h\lesssim 0.1\)) and
planet masses.  A real PPD may have significant surface density
variation that leads to a different ratio $T_{e,1}/T_{e,2}$.  Here we
study the effects of different $T_{e,1}/T_{e,2}$ on the equilibrium
values of \(e_i\) and whether such a change could lead to apsidal
alignment.

We explore this possibility in Figures \ref{fig:eqecc} and
\ref{fig:eqDpom} by varying \(T_{e,1}/T_{e,2}\) between \(0.2\) and
\(10\), regardless of the mass ratio.  The migration timescale is
still set to \(\abs{T_{m,i}}=T_{e,i}/(3.46 h^2)\).  We can see that
for comparable mass planets with \(q=0.5\), \(1\), and \(2\), varying
the ratio \(T_{e,1}/T_{e,2}\) around \(1/q\) modifies the final
equilibrium eccentricities by a roughly similar factor.  The dashed
lines of Fig.~\ref{fig:eqecc} show the analytic results from solving
equations~\eqref{dote1} -- \eqref{dotdpom} and \eqref{doteta}; these
results agree with our time dependent integrations.

For integrations where the
migration direction is the opposite of what it would be if
$T_{e,1}/T_{e,2} = 1/q = m_2/m_1$ (i.e., $T_{e,1}/T_{e,2} <1$ for
$q=0.5$), the analytical results systematically overestimate the final
eccentricities for $T_{e,1}/T_{e,2} <1$, and underestimate them for
$T_{e,1}/T_{e,2} > 1$.  The eccentricity ratio \(e_1/e_2\) is
unchanged, yet \(e_1\) and \(e_2\) are larger for more extreme values
of \(T_{e,1}/T_{e,2}\).  The corresponding values for \(\Delta\varpi\)
are shown in Fig.~\ref{fig:eqDpom}. In all cases, both the analytic
equilibrium equations and the numerical integrations yield
$\Delta\varpi\simeq 180^\circ$.

The peaked shape of the dashed lines in Figures
\ref{fig:standardeqecc} and \ref{fig:eqecc} can be explained as
follows.  As \(T_{m,1}/T_{m,2}=T_{e,1}/T_{e,2}\) approaches unity, the
effective migration timescale \(T_m\) approaches infinity.  Equation
\eqref{doteta} therefore implies that the planets' eccentricities
approach zero.  We note that the equilibrium solutions are not
continuous across $q=1$ (Fig.~\ref{fig:standard}) and
$T_{e,1}/T_{e,2}=1$ (Figures \ref{fig:eqecc} and \ref{fig:eqDpom}),
which is where we reverse the migration direction to ensure that it is
convergent.

In summary, we find that varying the eccentricity damping ratio cannot
account for the observed apsidal alignment in some MMR systems (such
as K2-19).  Before proposing a possible solution in Section
\ref{sec:orgd5c121f}, we consider in Section~\ref{sec:org4c72d92} the
simpler problem of a test mass orbiting near a MMR with a finite-mass
planet in order to gain some insight on the dynamics of apsidal
angles.

\section{MMR  in the Restricted 3-body Problem}
\label{sec:org4c72d92}

\begin{figure*} \centering
  \includegraphics[width=0.7\textwidth]{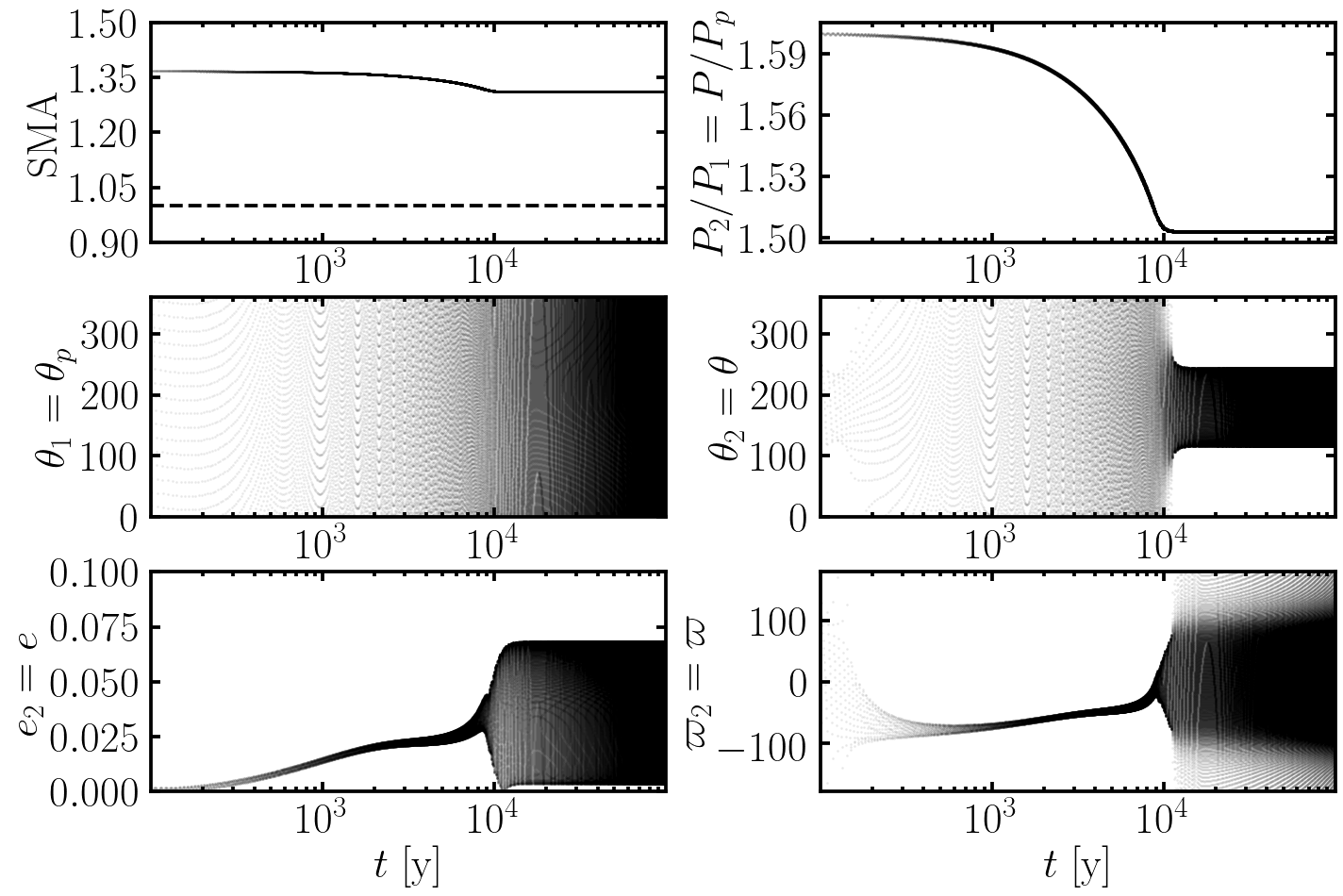}
  \caption{The capture process of a test particle into the 2:3 MMR
    with a massive planet with $\mu_p=10^{-4}$, $a_p=1$ au and
    eccentricity $e_p=0.04$.  We have set $h=0.03$ (which gives
    $e_{\rm eq}=0.028$) and $T_e = 1000P_p$.  After the particle is
    captured into resonance, both $\theta_2$ and $e$ librate with
    large amplitude, and $\varpi$ circulates.}
  \label{fig:tp-circ}
\end{figure*}

\begin{figure*}
  \centering
  \includegraphics[width=0.7\textwidth]{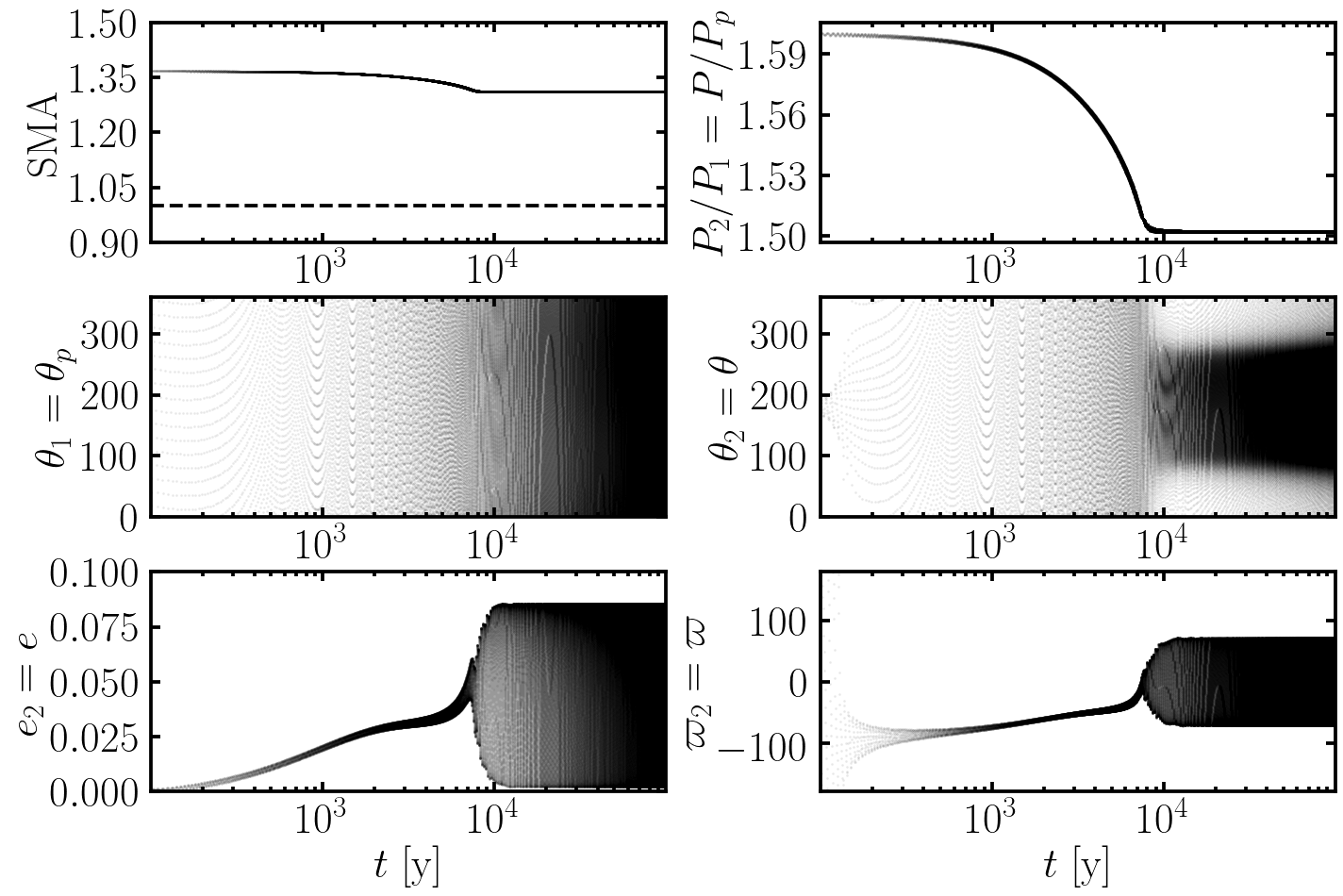}
  \caption{Same as Fig.~\ref{fig:tp-circ}, except $e_p=0.054$.
    Although $P_2/P_1$ settles down to a value very close to 3/2, the
    resonance angle $\theta_2$ circulates throughout the evolution.
    As in Fig.~\ref{fig:tp-circ}, the test particle eccentricity
    librates with large amplitude, but now its apsidal angle becomes
    aligned with $\varpi_p\equiv 0^\circ$.}
  \label{fig:tp-align}
\end{figure*}

\begin{figure} \centering
  \includegraphics[width=0.4\textwidth]{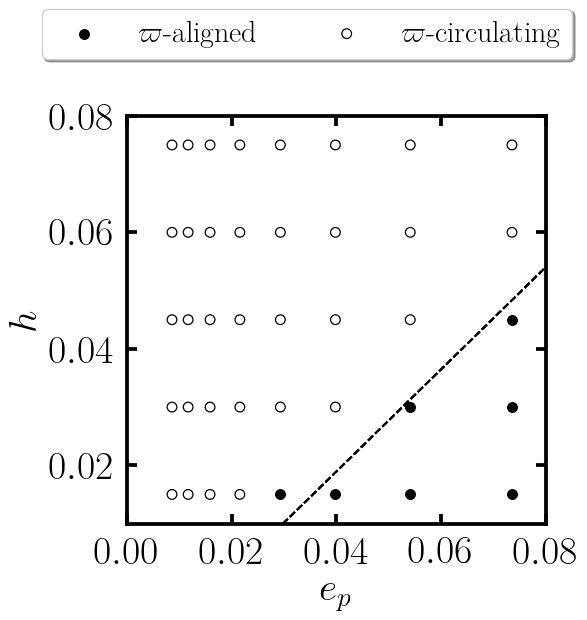}
  \caption{The behavior of the final $\Delta\varpi$ for a range of
    values for $h$ and $e_p$ when an exterior test mass is captured
    into the 2:3 MMR with a planet of mass $\mu_p=10^{-4}$. For a
    given $h$, the system becomes apsidally aligned for sufficiently
    large $e_p$ (see Fig.~\ref{fig:tp-align} for an example).  The
    dashed line indicates our analytical approximation for the
    boundary between the $\varpi$-aligned and $\varpi$-circulating
    regions.}
  \label{fig:tp-grid-ext}
\end{figure}

The dynamics of first-order MMRs with comparable mass planets is
complicated by the presence of two critical arguments in the
Hamiltonian, \(\theta_1\) and \(\theta_2\).  By assuming that one of
the planets is a test particle, we may ignore the dynamical evolution
of the other planet.  To emphasize the fact that we are formally
transitioning to a different problem, we adopt the following notation:
The subscript \(p\) will denote the quantities associated with the
massive planet, while no subscript will indicate those associated with
the test particle.  Neglecting additive constants, the Hamiltonian for
an exterior test particle ($m_2=0$) is given by
\begin{align}
  \label{eq:tpext}
  H = &-\frac{GM}{2a} -\frac{Gm_p}{a_p}[f_1e_p\cos\theta_p + f_2e\cos\theta \nonumber\\
  &+f_3e^2 + f_4ee_p\cos(\varpi_1-\varpi_2)].
\end{align}
We assume that no dissipative force acts on the massive planet
while implementing the same force as in
Section~\ref{sec:org985dec7} for the test particle.

\subsection{Circular massive planet: $e_p=0$}
\label{sec:org6c9c017}
When a test particle is caught into a stable resonance, its
eccentricity grows and saturates at a finite value which depends on
the ratio \(T_{e}/T_{m}\)
\citep[e.g.][]{goldreich_overstable_2014,xu_migration_2018}.
For an exterior test particle,
\begin{equation}
\label{eq:eeqext}
  e_{\rm eq} = \sqrt{\frac{T_e}{2jT_m}} = h\sqrt{\frac{1.73}{j}}.
\end{equation}
If $e_p=0$, the system has no preferred direction, and so
$\varpi$ must circulate.

\subsection{Eccentric massive planet}
\label{sec:orgc4cb86e}

When the massive planet has a moderate eccentricity, the qualitative
features of the $e_p=0$ result are preserved as long as
\(e_p \lesssim e_{\rm eq}\). In Fig.~\ref{fig:tp-circ}, we show the
capture process for a system with \(e_p = 0.04\) and $h=0.03$ (thus
$e_{\rm eq}=0.028$).  As we can see, the particle is still captured
into the 2:3 resonance and \(\theta_2\) librates around
180\(^\circ\). The eccentricity \(e\) librates around its equilibrium
value with large amplitude, and the longitude of perihelion,
\(\varpi\), still circulates.

Fig.~\ref{fig:tp-align} shows the capture process for $e_p=0.054$
(and the same $e_{\rm eq}=0.028$).  We see that, for
\(e_p \gtrsim e_{\rm eq}\), the test particle's migration halts near
the nominal resonance location of \(P_2/P_1=1.5\) while both
\(\theta_1\) and \(\theta_2\) continue to circulate. The particle's
eccentricity librates with slightly larger amplitude than in 
Fig.~\ref{fig:tp-circ}. Eventually, the system becomes apsidally aligned.

In Fig.~\ref{fig:tp-grid-ext}, we summarize the behavior of the final
\(\Delta\varpi\) for an exterior test particle
captured into the 2:3 MMR with a massive planet,
for different values of \(e_p\) and \(h\) (and subsequently \(e_{\rm
eq}\)). Generally, for \(e_{p}\gtrsim e_{\rm eq}\), the system
becomes apsidally aligned.

\section{Eccentricity Driving by Disc and Apsidal Alignment in MMR}
\label{sec:orgd5c121f}
We have seen in Section~\ref{sec:org985dec7} that whenever two
comparable-mass planets are captured into the $\theta_1$ and
$\theta_2$ resonances, the system always has
$\Delta\varpi\approx180^\circ$ because $\theta_1$ and $\theta_2$
settle down to $180^\circ$ and $0^\circ$, respectively.  The apsidally
aligned K2-19 system (see Section~\ref{sec:org493ee54}) therefore
poses a problem for the standard migration-driven MMR capture model.
In order to match this observation, either \(\theta_1\), \(\theta_2\),
or both angles must circulate.

In Section~\ref{sec:org4c72d92}, we have seen that apsidal alignment
arises whenever the massive planet has an eccentricity larger than the
equilibrium eccentricity of the test particle in resonance.
Guided by this result, in this section, we
examine the possibility that planet-disc interaction drives planet's
eccentricity to a finite value and explore its consequences for
the apsidal angles in MMR capture.
In addition, we reformulate the 2-planet Hamiltonian into a
single-degree of freedom system; this allows us to identify
the key dynamical processes that lead to apsidal alignment.

\subsection{Effects of disc eccentricity driving on MMR}
\label{sec:orgdf70eb9}
\begin{figure*}
  \centering
  \includegraphics[width=0.7\textwidth]{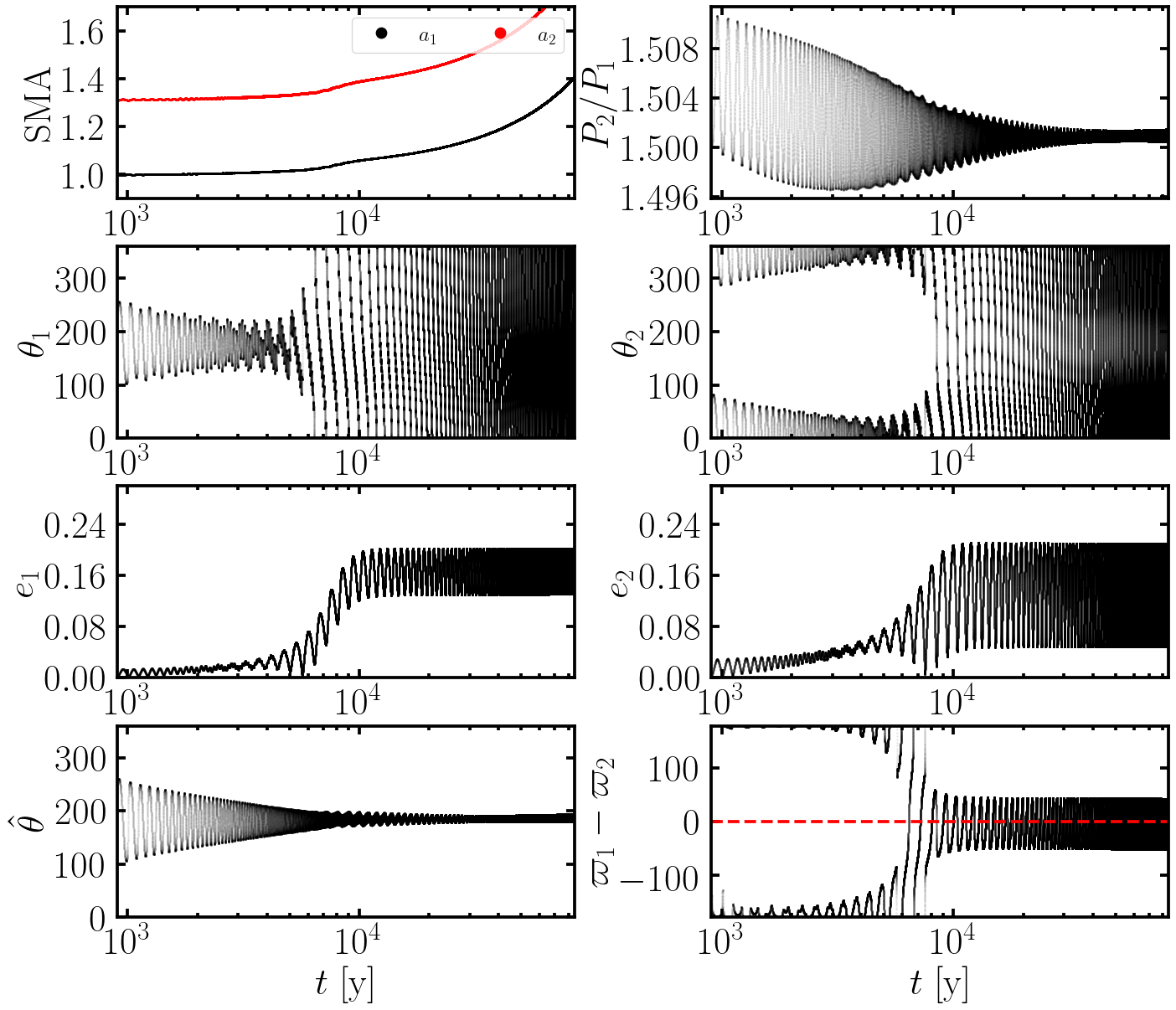}
  \caption{Capture of comparable mass ($q=m_1/m_2=2$) planets into the
    3:2 MMR with $e_{1,d}=0.2$, $e_{2,d}=0$ (see equation
    \ref{eq:dotedriving})and $h=0.03$. All other initial conditions
    are the same as in Fig.~\ref{fig:standardex}, except for the
    initial period ratio, which we set to the nominal resonance
    location, $P_2/P_1=1.5$, so that the system is very quickly caught
    into the $\theta_1$ and $\theta_2$ resonances.  After about
    10~kyr, the system escapes the circular resonances, indicated by
    the circulation of $\theta_1$ and $\theta_2$. At this point, the
    planets becomes apsidally aligned and $\Delta\varpi$ librates
    around $0^\circ$.}
  \label{fig:drivingex}
\end{figure*}

\begin{figure*}
  \centering
  \includegraphics[width=0.7\textwidth]{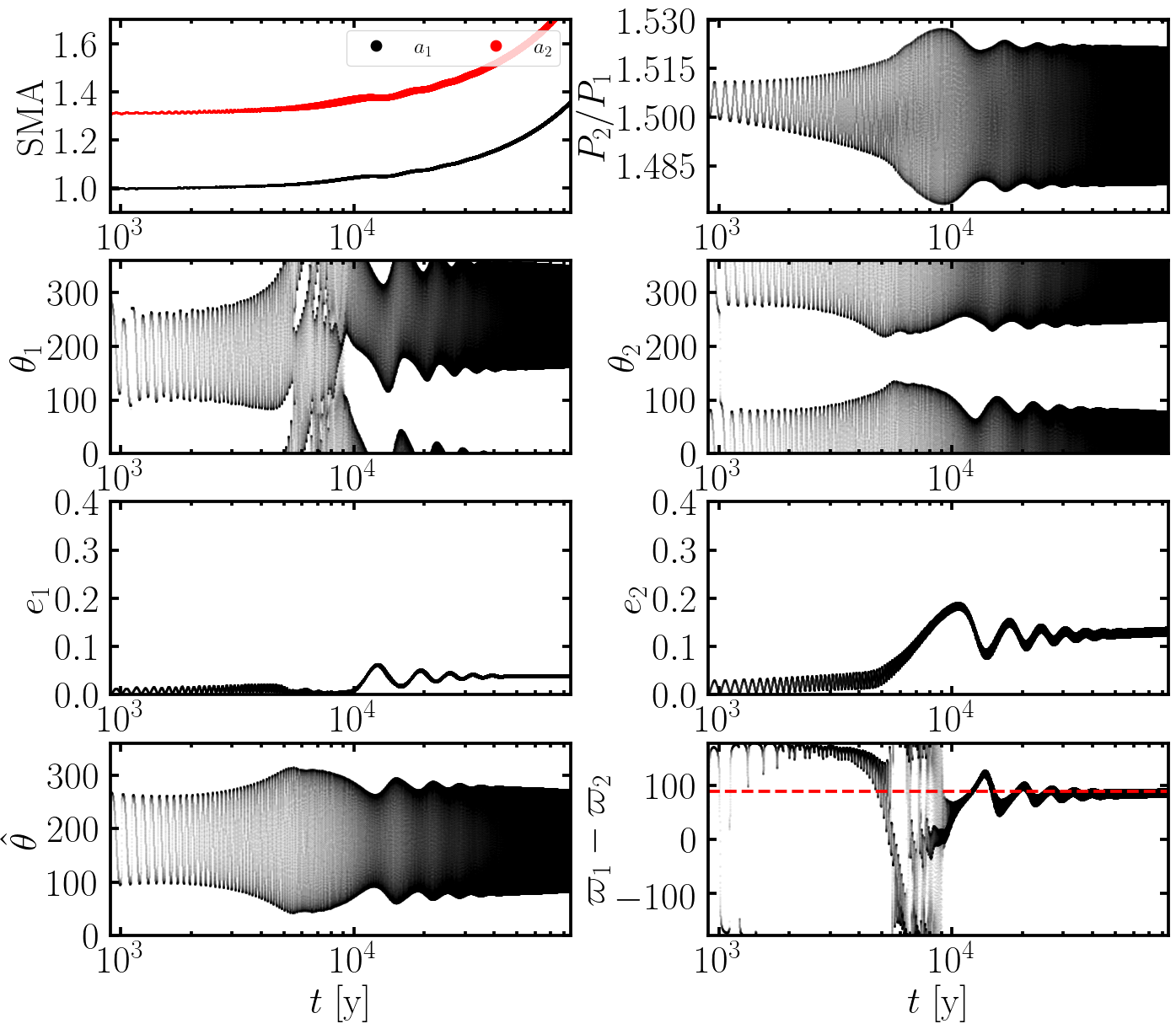}
  \caption{Same as Fig.~\ref{fig:drivingex}, but with $e_{1,d}=0$
    and $e_{2,d}=0.2$.  After about 10~kyr, the system escapes the
    $\theta_1$ resonance, indicated by its shift in libration center
    to $270^\circ$ rather than $180^\circ$.  We see that $\theta_2$
    still librates around $0^\circ$, and so the apsidal angles of the
    planets reach $\Delta\varpi=90^\circ$.}
  \label{fig:perpex}
\end{figure*}
As noted in Section~\ref{sec:org493ee54}, under appropriate
conditions, planet-disc interactions can \emph{increase} a planet's
eccentricity rather than damp it
\citep[e.g.][]{goldreich03_eccen_evolut_planet_gaseous_disks,teyssandier17_secul_evolut_eccen_protop_discs,ragusa17_eccen_evolut_durin_planet_disc_inter}.
A recent study demonstrates that a super-earth-sized luminous
protoplanet can attain an eccentricity larger than the disc aspect
ratio \citep{romero21_eccen_drivin_pebbl_accret_low_mass_planet}.

The planets K2-19b and c are moderately eccentric, with
\(e_{b}\approx e_c\approx 0.2\) \citep{petigura_k2-19b_2020}.
\citet{petit_resonance_2020} suggest that the apsidal alignment in
this system could be caused by eccentricity driving to a common value.
To mimic the effect of eccentricity driving by the disk, we modify the
eccentricity damping term in equation~\eqref{eq:disforce} to
\begin{equation}
  \label{eq:dotedriving}
  \frac{\dot e_i}{e_i} = -\frac{(e_i-e_{i,d})}{T_{e,i}},
\end{equation}
\noindent so that planet \(m_i\) is driven toward \(e_{i,d}\) on the
timescale \(T_{e,i}\).  

In Fig.~\ref{fig:drivingex}, we
show the result of MMR capture for a system with
\(e_{1,d}=0.2\), \(e_{2,d}=0.0\), and mass ratio $q=m_1/m_2=2$.
We initialize the system with \(e_{1}=e_2=0.001\) at the nominal
resonance location, \(P_{2}/P_1 = 1.5\). The planets are caught in the
\(\theta_1\) and \(\theta_2\) resonances for 10,000 years, after which
the planets escape the resonance and the angles circulate.  At this
point, both planets' eccentricities are excited to about
\(e_i\approx 0.2\) and the planets become apsidally aligned as
\(\Delta\varpi\) librates around \(0^\circ\) with a large amplitude.
Despite the circulation of both resonance angles, the period ratio
remains locked very close to the nominal resonance value
(\(P_2/P_1= 1.5\)). The system is caught in a different type of
resonance which we will study in the following subsection.

On the other hand, for \(e_{2,d}=0.2\) and \(e_{1,d}=0\), the system
displays different resonance capture behavior.  We show the result for
this case in Fig.~\ref{fig:perpex}. We see that the angle
\(\theta_2\) librates with a large amplitude around its resonant value
of \(0^\circ\), whereas \(\theta_1\) librates around \(270^\circ\)
rather than \(180^\circ\).  As a result, \(\Delta\varpi\) approaches
\(90^\circ\), i.e. the planets' perihelia are now perpendicular to
each other.

\subsection{Reducing the Hamiltonian}
\label{sec:org8116cb5}
\begin{figure}
  \centering
  \includegraphics[width=0.45\textwidth]{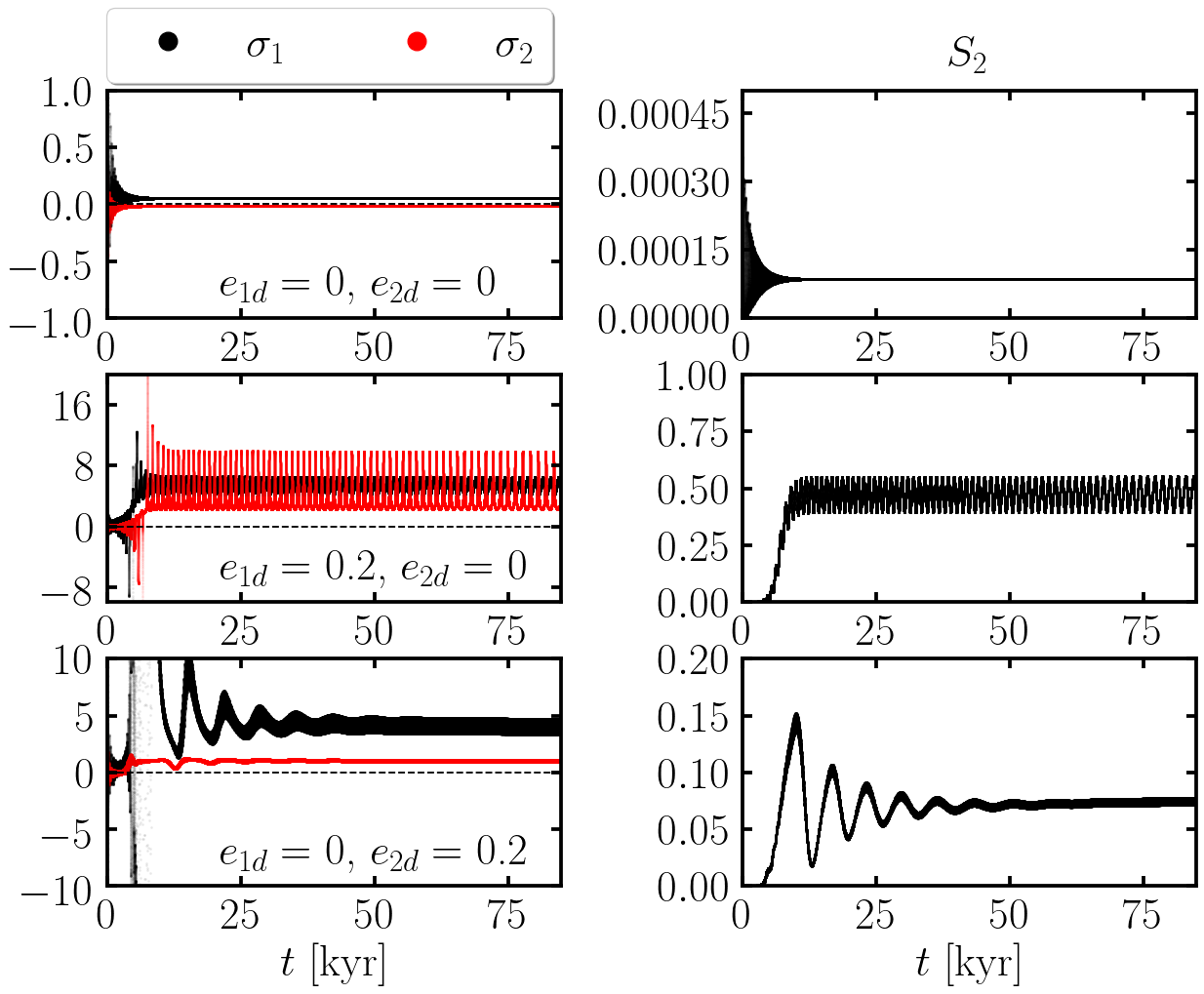}
  \caption{
    The quantities $\sigma_1$ and 
    $\sigma_2$ (left) and the quantity $S_2$ (right,
    see equations~\ref{eq:s2}, \ref{eq:sS1} and \ref{eq:sS2})
    for the three different combinations of $e_{1d}$ and $e_{2d}$
    corresponding to the systems depicted in Figures \ref{fig:standardex}, \ref{fig:drivingex}, and
    \ref{fig:perpex}. Each row corresponds to one of the three
    different modes of resonance identified in this paper,
    $\Delta\varpi=180^\circ$, $\Delta\varpi=0^\circ$, and
    $\Delta\varpi=90^\circ$, respectively.
    For $\Delta\varpi=180^\circ$ (upper row),
    $\sigma_1$, $\sigma_2$, and $S_2$ are conserved near
    zero. For the other two cases, $S_2$ transitions to larger values
    near $\sim 0.5$ (middle) and $\sim 0.1$ (bottom) as the
    eccentricity reaches an equilibrium.  Both $\sigma_1$
    and $\sigma_2$ are excited to factors of a few in the
    apsidally aligned case (middle row), while only $\sigma_1$ is excited for the
    $\Delta\varpi=90^\circ$ case (bottom row).  For $\Delta\varpi=0^\circ$ and
    $\Delta\varpi=90^\circ$ cases, the eccentricities oscillate in such a
    way to conserve $S_2$ according to equation (\ref{eq:S2eq}).}
  \label{fig:S2cons}
\end{figure}

\begin{figure}
  \centering
  \includegraphics[width=0.45\textwidth]{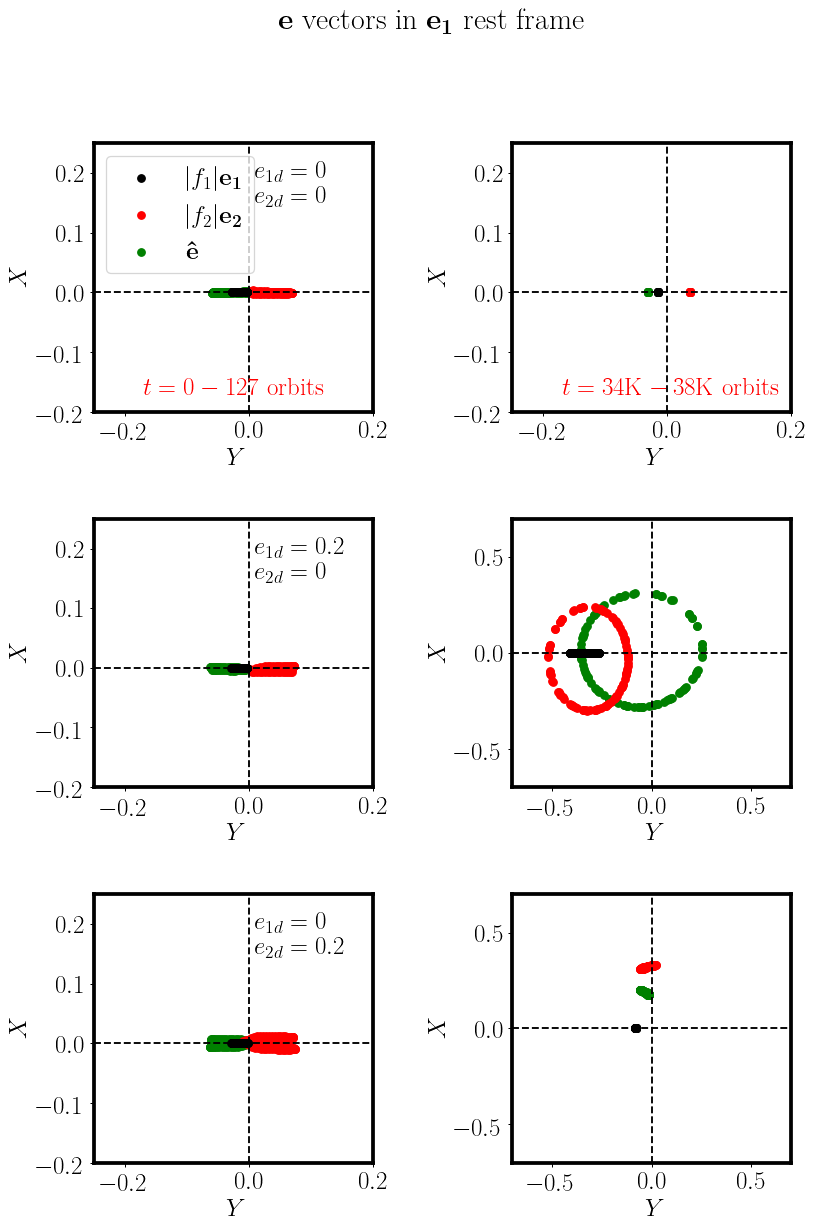}
  \caption{The eccentricity vectors $\mathbf{e}_1$, $\mathbf{e}_2$,
    and $\mathbf{\hat e}$ in the reference frame rotating with $\mathbf{e}_1$. As
    in Fig.~\ref{fig:S2cons}, each row represents the different combination of
    $e_{1d}$ and $e_{2d}$ corresponding to the three different modes
    of resonance. The left column shows the initial conditions of the
    resonance, while the right column shows the evolution at late
    times (at $t=[34-38]\times10^3$ orbits). The top row ($\Delta\varpi=180^\circ$) exhibits little
    qualitative change between the initial and late times besides the libration amplitudes shrinking to
    zero.  The second row ($\Delta\varpi=0^\circ$) shows
    $\mathbf{e}_2$ circulating around $\mathbf{e}_1$ strictly
    contained to the second and third quadrants. The last row
    ($\Delta\varpi=90^\circ$) shows $\mathbf{e}_1$ and $\mathbf{e}_2$
    transitioning into a perpendicular arrangement. Meanwhile, the
    $\mathbf{\hat e}$ vector circulates outside of both $\mathbf{e}_1$
    and $\mathbf{e}_2$ in the second row, while remaining aligned with
    $\mathbf{e}_2$ in the bottom row.}
  \label{fig:relgeom}
\end{figure}

To understand the results of Section~\ref{sec:orgdf70eb9}, we carry
out an analysis of the MMR Hamiltonian (equation~\ref{hamiltonian}).
This helps to illustrate the underlying dynamics behind the capture
process in Fig.~\ref{fig:drivingex} which leads to apsidal
alignment.  It can be shown that \(\theta_1\) and \(\theta_2\) are
actually subresonances of a resonance \(\hat\theta\), which arises
after transforming the system's Hamiltonian so that it has only a
single degree of freedom
\citep[][]{henrard86_reduc_trans_apocen_librat,deck13_first_order_reson_overl_stabil}.

If we assume that the secular behavior of the semi-major axis ratio
\(\alpha\) is stationary or varying adiabatically, we may transform the
resonant Hamiltonian \(H_{\rm Kep} + H_{\rm res}\) in equation
\eqref{hamiltonian} into the form
\begin{equation}
  \label{hhat}
  \hat H(\hat R,\hat\theta) = -3(\delta+1) \hat R + \hat R^2 - 2\sqrt{2\hat R} \cos\hat\theta
\end{equation}
\noindent through a series of rotations in phase space.  For
the details of these transformations, see Appendix
\ref{sec:org2b22ae8}.
The Hamiltonian in equation~\eqref{hhat} is well studied
in the literature
\citep[e.g.,][]{murray_solar_2000}.
The parameter \(\delta\) quantifies
the system's depth into resonance.  We do not include \(H_{\rm sec}\) in
this analysis because it is second order in eccentricities.

In equation~\eqref{hhat}, the action $\hat R$ is a function of both
\(e_1\) and \(e_2\).  Define
\(\v{\hat e} = \abs{f_1}\v e_1 - \abs{f_2}\v e_2\) and
\(\hat e = \abs{\v{\hat e}}\), where \(\v e_i\) is the eccentricity (Runge-Lenz)
vector of each planet. The action \(\hat R\) takes the form
\(\hat R \propto \tilde \mu \hat e^2\), where \(\tilde\mu=\tilde m/M\)
and \(\tilde m= m_1m_2/(m_1+m_2)\).  The coordinate angle, \(\hat\theta\), is given by
\begin{align}
\label{eq:hattheta}
  \tan\hat{\theta}  = \frac{f_1 e_1\sin\theta_1
  + f_2e_2\sin\theta_2}{f_1e_1\cos\theta_1 + f_2e_2\cos\theta_2}.
\end{align}
\noindent
This angle is the same one that \citet{petit_resonance_2020}
found to librate in the K2-19 system.

\subsection{Three modes of resonance}
\label{sec:org28f6acc}
The one-degree-of-freedom Hamiltonian (equation~\ref{hhat})
admits the following conserved quantity:
\begin{align}
  \label{eq:s2}
  S_2 = q\sqrt{\alpha}f_2^2e_1^2
+2\abs{f_1f_2}e_1e_2\cos(\varpi_1-\varpi_2) + \frac{f_1^2}{q\sqrt\alpha}e_2^2.
\end{align}

\noindent
By enforcing \(dS_2/dt = 0\) together with the assumption \(d\alpha/dt
= 0\), we arrive at the following equilibrium condition:
\begin{align}
  \label{eq:S2eq}
  \frac{dS_2}{dt} \propto e_1^2\left(\frac{e_1-e_{1\rm d}}{T_{e,1}}\right)\abs*{\frac{f_2}{f_1}}
  \sigma_1
  + e_2^2\left(\frac{e_2-e_{2\rm d}}{T_{e,2}}\right)
  \sigma_2
  = 0,
\end{align}
\noindent
where
\begin{align}
  \label{eq:sS1}
  \sigma_1=&\left[
                  q^2\alpha\abs*{\frac{f_2}{f_1}}
                  + \frac{e_2}{e_1}q\sqrt{\alpha}\cos(\varpi_1-\varpi_2)
                  \right],\\
  \label{eq:sS2}
  \sigma_2=&\left[
                  \abs*{\frac{f_2}{f_1}} q\sqrt{\alpha}
                  \frac{e_1}{e_2}\cos(\varpi_1-\varpi_2) + 1
                  \right].
\end{align}

The systems depicted in Figures \ref{fig:standardex},
\ref{fig:drivingex}, and \ref{fig:perpex} are representative of three
different modes of resonance, ones with
\(\Delta\varpi\simeq180^\circ\), \(\Delta\varpi\simeq0^\circ\), and
\(\Delta\varpi\simeq90^\circ\), respectively. These correspond to
three different behaviors of the quantities \(\sigma_1\) and
\(\sigma_2\) while in resonance under the influence of eccentricity
forcing.  In Fig.~\ref{fig:S2cons}, we show \(\sigma_1\),
\(\sigma_2\), and \(S_2\) for these systems. The top row is
for the standard eccentricity damping case where \(e_{1d}=e_{2d}=0\).
Once the system equilibrates, \(S_2\approx 10^{-4}\) is well conserved
(top left) and small. Both \(\sigma_1\) and \(\sigma_2\) are also
close to zero. From equations~\eqref{eq:sS1} and \eqref{eq:sS2}, we
see this corresponds to \(e_2/e_1 \sim q\), as we found in Section
\ref{sec:org985dec7}.  The second row of Fig.~\ref{fig:S2cons}
corresponds to the system shown in Fig.~\ref{fig:drivingex},
where \(e_{1d}=0.2\) and \(e_{2d}=0\).  At early times, while the
system is still caught in the \(\theta_1\) and \(\theta_2\)
resonances, \(\sigma_1\), \(\sigma_2\), and \(S_2\) are small.  Once
the \(\theta_1\) and $\theta_2$ resonances are broken, and only \(\hat\theta\)
librates, $\sigma_1$, $\sigma_2$ and $S_2$ are excited to larger
values.  The quantities \(\sigma_1\) and \(\sigma_2\) undergo large
periodic oscillations away from zero, while \(S_2\) grows and then
stabilizes at its new equilibrium value, \(S_2\approx 0.5\). The
planets' eccentricities oscillate in such a way as to conserve
\(S_2\).  The bottom row of Fig.~\ref{fig:S2cons} corresponds to the
system in Fig.~\ref{fig:perpex}, where \(e_{2d}=0.2\) and
\(e_{1d}=0\), where the planets' perihelia are perpendicular.  For
this case, \(\sigma_2\) is conserved close to 0, while \(\sigma_1\)
grows to a magnitude similar to its value in the apsidally aligned
case.

In Fig.~\ref{fig:relgeom}, we plot the eccentricity
vectors \(\mathbf{e}_{1}\), \(\mathbf{e}_{2}\), and \(\mathbf{\hat e}\)
in the reference frame rotating with \(\mathbf{e}_1\).
The three systems begin with the same configuration,
caught in the \(\theta_1\) and \(\theta_2\) resonances;
the vectors \(\mathbf{e}_1\) and \(\mathbf{e}_2\) are anti-aligned,
while \(\mathbf{\hat e}\) is aligned with \(\mathbf{e}_1\).
At later times, the system without eccentricity driving
remains in this configuration (top row).
The second and third rows exhibit the new resonance
behaviors described above.  For the apsidally aligned case (middle row),
\(\mathbf{e_2}\) circulates in a pattern strictly constrained to the
second and third quadrants, and \(\mathbf{\hat e}\) circulates around
the other two vectors.  In the bottom row, \(\mathbf{e}_1\) and
\(\mathbf{e}_2\) are perpendicular to each other and \(\mathbf{\hat e}\)
is aligned with \(\mathbf{e}_2\).

\subsection{Dependence on $e_{\rm 1,d}$ and $e_{\rm 2,d}$}
\label{sec:orgda57f89}
\begin{figure} \centering
\includegraphics[width=0.4\textwidth]{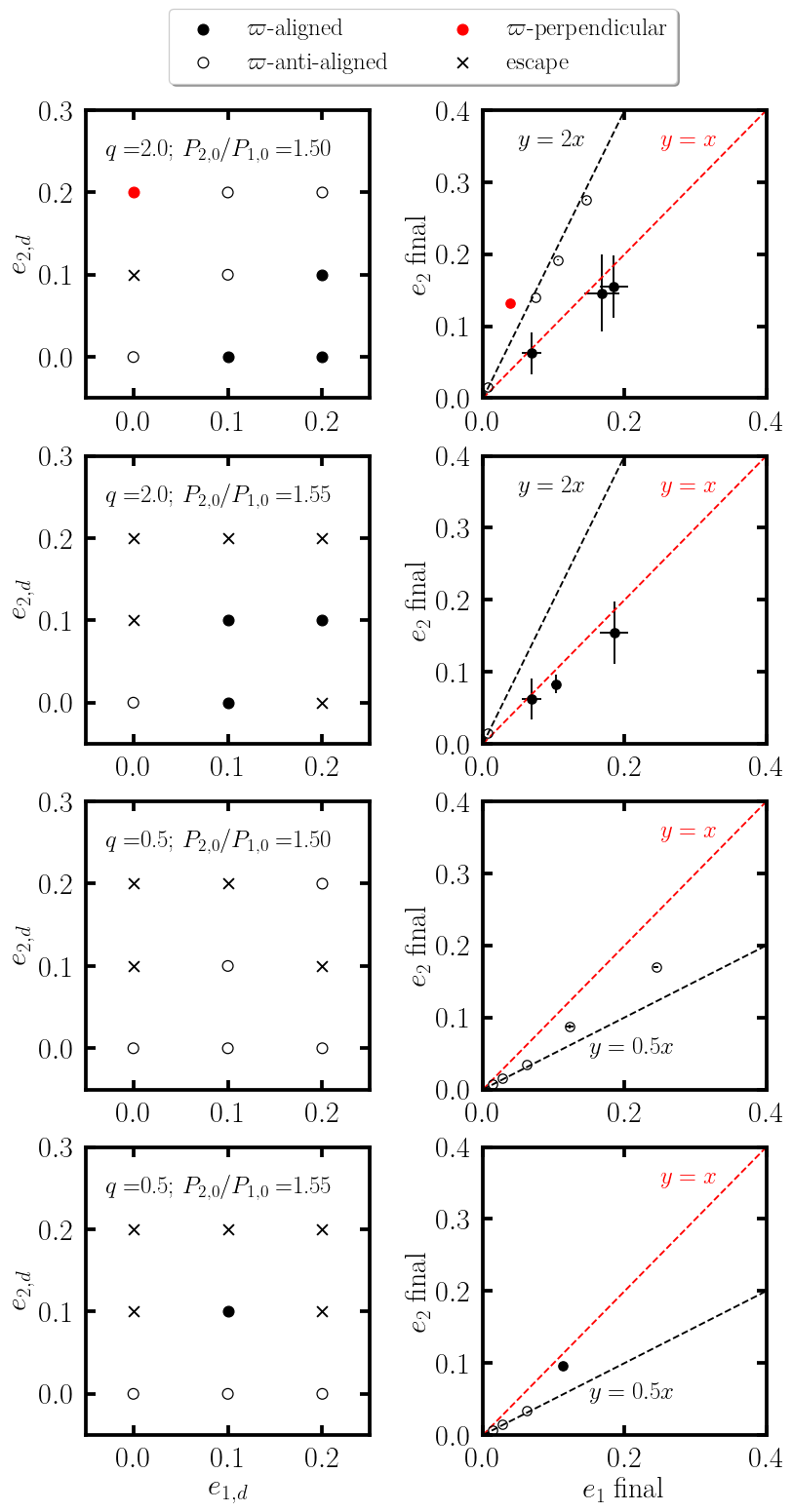}
\caption{The resonance architecture for systems spanning a grid of
  driving eccentricities $e_{1,d}, e_{2,d}$ (left) and the numerically
  averaged final eccentricities (right).  We hold $h=0.03$ and
  $T_{e,0}=1000$~yrs, while varying $e_{1,d}$ and $e_{2,d}$.  Systems
  which are not captured or become unstable and escape resonance are
  denoted by black ``x''-marks. Roughly, for $e_{1d}\gtrsim e_{2d}$,
  the system becomes aligned.  For $e_{2d}\gtrsim e_{1d}$, the system
  becomes perpendicular ($\Delta\varpi=90^\circ$) for large values of
  $e_{2d}$. The other systems remain trapped in both of the $\theta_1$
  and $\theta_2$ resonances.  The time-averaged eccentricities of the
  apsidally aligned systems fall just below the line $e_2=e_1$. The
  error bars are their standard deviations. The single perpendicular
  system (see the top left panel) falls just above the line
  $e_2=2e_1=qe_1$, while the anti-aligned systems fall just below
  it. Both have small librations compared to the aligned case.}
    \label{fig:Rhat-grid} 
  \end{figure}

\begin{figure*}
  \centering
  \includegraphics[width=0.7\textwidth]{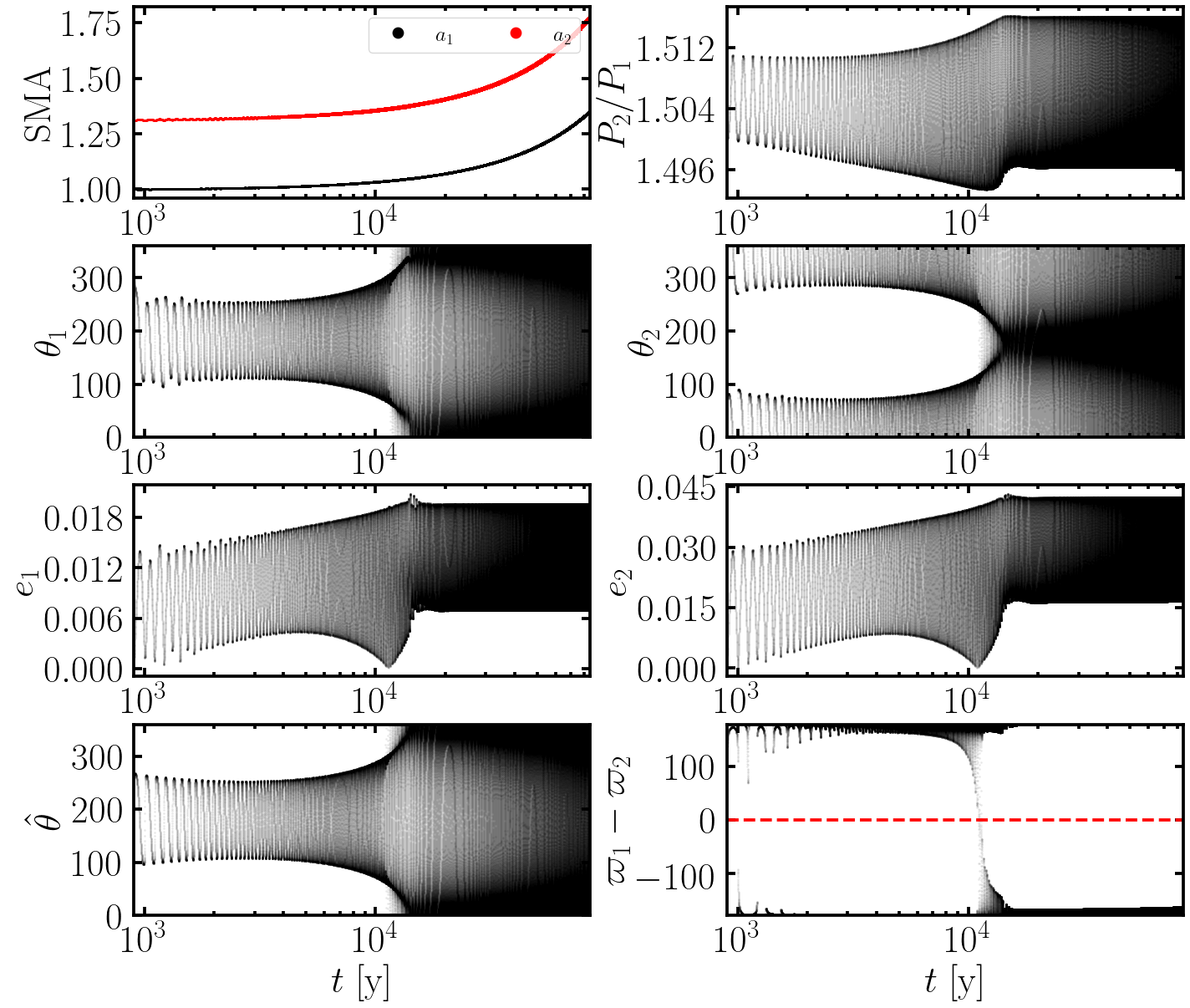}
  \caption{ Same as Fig.~\ref{fig:drivingex}, but with $e_{1d}=0$,
    $e_{2d}=0.1$.  This system corresponds to the $\mathtt{x}$ marker
    in the top left panel of Fig.~\ref{fig:Rhat-grid}. The system
    starts off in resonance, with $\theta_1$, $\theta_2$ and
    $\hat\theta$ librating. However, after $10^4$ years, the system
    breaks out of all three resonances. Nevertheless, the period ratio
    remains locked around $1.5$ with small librations. The
    eccentricities reach the equilibrium values with large librations,
    while the apsidal angle $\Delta\varpi$ transitions from
    slightly less than $180^\circ$ to slightly
    larger than $180^\circ$.}
  \label{fig:escapeex}
\end{figure*}

Now that we have identified these three resonant modes, here we
explore the \((e_{1d},e_{2d})\) parameter space for moderate values
between \(0\) and \(0.2\).  The top row of Fig.~\ref{fig:Rhat-grid}
shows the result for \(q=2\), \(h=0.03\), and initial period ratio
\(P_2/P_1=1.5\).  In the left panel, we summarize the resonant
behavior for each system on an $e_{1,d}$-$e_{2,d}$ grid.  We mark the
run which becomes unstable and escapes the resonance within the
timescale of our integration by an $\mathtt x$ marker. Roughly, for
\(e_{1d}\gtrsim e_{2d}\), the system becomes apsidally aligned
($\Delta\varpi=0^\circ$). For \(e_{2d} > e_{1d}\), one case exhibits
\(\Delta\varpi=90^\circ\), one escapes, and the others are apsidally
anti-aligned.  In the right panel of Fig.~\ref{fig:Rhat-grid}, we
plot the time-averaged final eccentricities of the planets. The points share
the same color-coding as in the left panel.  The eccentricities for
the aligned cases fall roughly along the line \(e_1=e_2\), which
reflects the fact that the angle \(\hat\theta\) does not depend on the
mass ratio \(q\) (equation~\ref{eq:hattheta}).  The perpendicular case ($\Delta\varpi=90^\circ$)
falls slightly above the line \(e_2/e_1=q\), while the anti-aligned
runs fall just under it.

The single \(\mathtt{x}\) marker in the top left panel of Fig.~\ref{fig:Rhat-grid} corresponds to a
run which is only temporarily caught into resonance.
We show the detailed evolution of this system
in Fig.~\ref{fig:escapeex}. Although  all
resonance angles circulate, the period ratio librates with a small
amplitude around \(P_2/P_1=1.5\). The planets remain in
an anti-aligned configuration throughout. Before the escape,
\(\Delta\varpi\lesssim180^\circ\), while after the escape,
\(\Delta\varpi\gtrsim180^\circ\).

\subsection{Initial period ratio}
\label{sec:org4f3aa47}
\begin{figure*}
  \centering
  \includegraphics[width=0.7\textwidth]{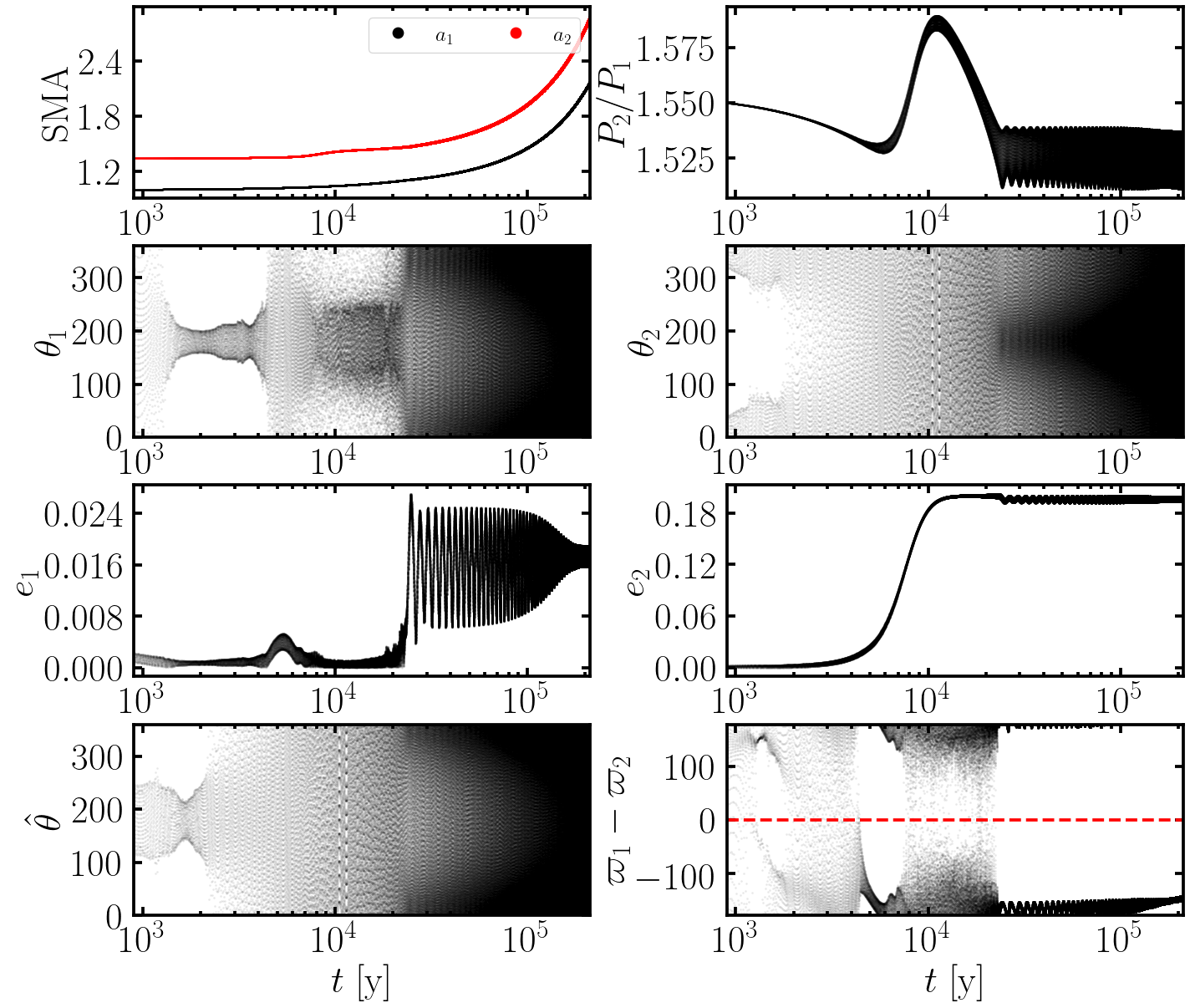}
  \caption{ Same as Fig.~\ref{fig:drivingex} (for $q=2$) but with
    $e_{1d}=0$ and $e_{2d}=0.2$ and an initial period ratio of
    $P_2/P_1=1.55$.  The system corresponds to one of the $\mathtt{x}$
    marks in the left, second row of Fig.~\ref{fig:Rhat-grid}.  The
    period ratio initially decreases as the planets migrate
    convergently. Around $t=10^4$ years, the period ratio increases,
    then again turns around and settles down into libration around
    1.525, while all resonant angles circulate.}
  \label{fig:escapeex1}
\end{figure*}

\begin{figure*}
  \centering
  \includegraphics[width=0.7\textwidth]{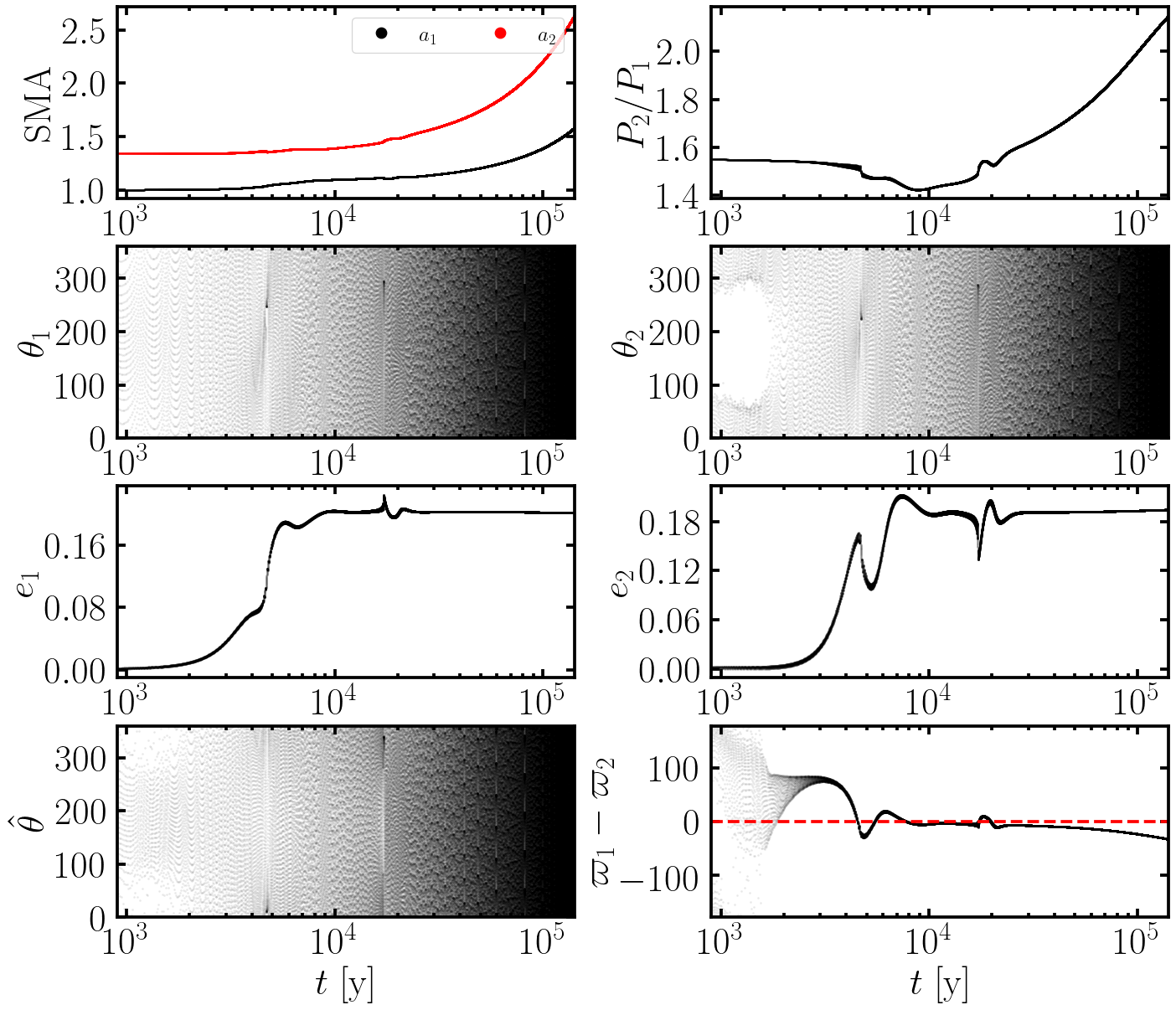}
  \caption{ Same as Fig.~\ref{fig:escapeex1}, but with $e_{1d}=0.2$
    and $e_{2d}=0.2$.  This corresponds to one of the $\mathtt{x}$
    marks in the left, second row of Fig.~\ref{fig:Rhat-grid}.  The
    system convergently migrates outward for some time, after which it
    crosses the resonance location without being captured (i.e., the
    kink between $t=3000$ and 4000 years).  The period ratio then
    turns around and misses the resonance again, after which it
    continues to grow for the rest of the integration.  In this
    example, as opposed to Fig.~\ref{fig:escapeex1}, the resonance
    angles never appear to librate.}
  \label{fig:escapeex2}
\end{figure*}
In the second row of Fig.~\ref{fig:Rhat-grid}, we have kept all of the
parameters from the first row constant, but shifted the initial
location of \(m_2\) so that \(P_2/P_1=1.55\). As we can see from the left panel,
many more systems fail to be permanently captured.

Fig.~\ref{fig:escapeex1} displays the result of a system
which fails to
be captured in the resonance. We see that
after a short period
of convergent migration, the outer planet is repelled
away from the resonance. Following an initial decrease,
the period ratio increases and then turns around
and levels off into a state which
resembles the late-time behavior in Fig.~\ref{fig:escapeex}.

Fig.~\ref{fig:escapeex2} depicts another example of failed resonance capture.  Initially, both planets undergo
convergent outward migration, but after some time, \(a_1\) begins to
increase.  Then, the planets skip the resonance as the period ratio
passes through \(P_2/P_1=1.5\) from above. It continues decreasing,
turns around, and then skips the resonance from below.

\subsection{Mass ratio}
\label{sec:org026e50b}
\begin{figure*}
  \centering
  \includegraphics[width=0.7\textwidth]{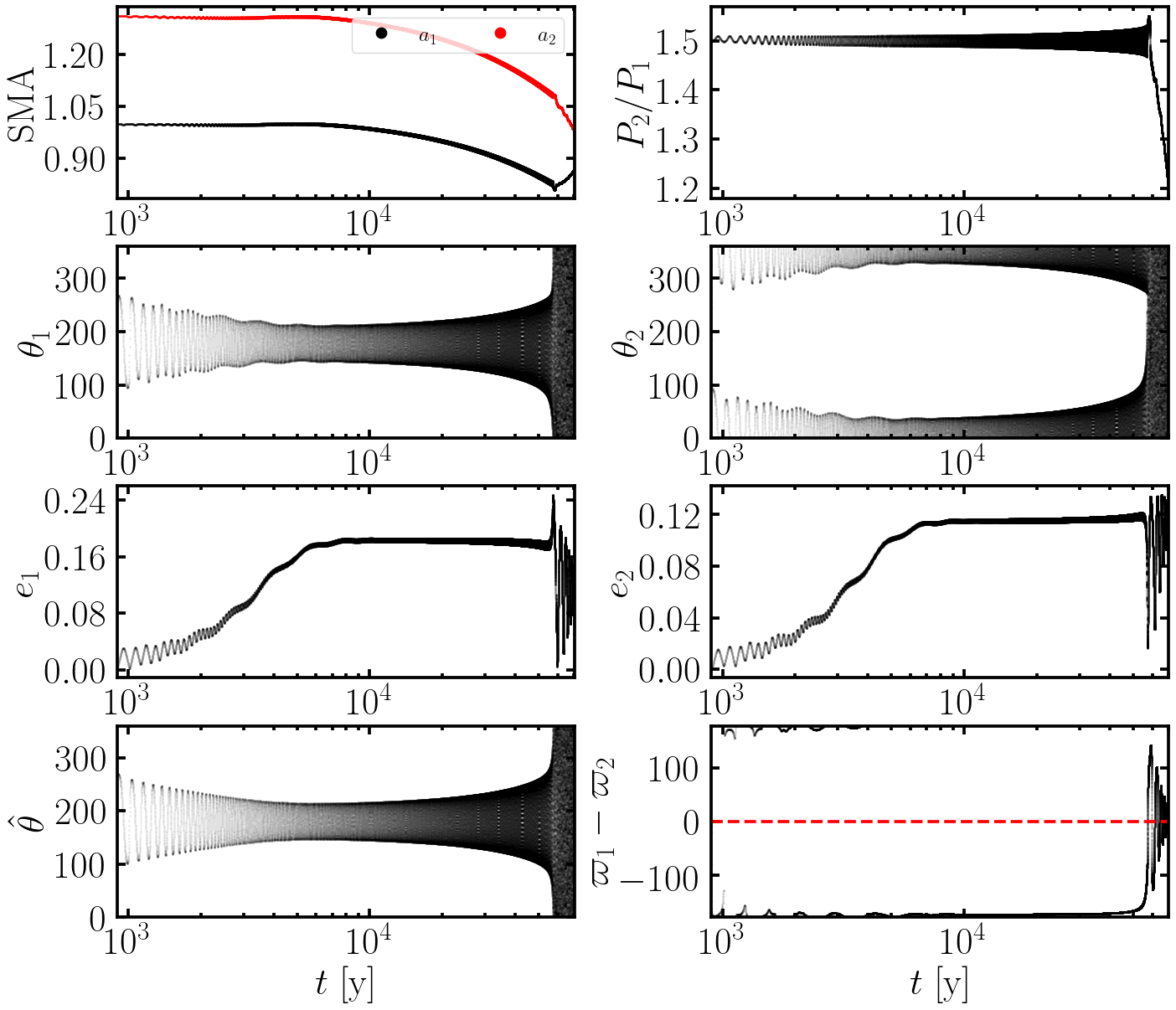}
  \caption{Same as Fig.~\ref{fig:escapeex1}, but with $q=0.5$,
    $e_{1d}=0.2$, $e_{2d}=0.1$ and an initial period ratio
    $P_2/P_1=1.5$.  This corresponds to one of the $\mathtt{x}$ marks
    in the left, third row of Fig.~\ref{fig:Rhat-grid}.  The planets
    are caught into all three resonances for most of the integration.
    The period ratio is likewise locked near 1.5 with libration that
    grows slightly over time. Then the system exits all three
    resonances at the same time, after which $m_1$ launches outwards
    leading to eventual orbit-crossing.}
  \label{fig:escapeex3}
\end{figure*}

Now we turn to the effect of mass ratio on the resonance capture.
Rows three and four of Fig.~\ref{fig:Rhat-grid} summarize the
results of running identical integrations to the first two rows but
setting \(q=0.5\), reversing the migration direction, and modifying
the dissipation timescales appropriately.  Only one system out of the
18 in the last two rows of Fig.~\ref{fig:Rhat-grid} becomes apsidally aligned, while
for all cases with \(e_{2,d}=0\), the planets are anti-aligned.  The
top-left-most points (i.e. \(e_{2,d}>e_{1,d}\)) all escape the resonance
whenever \(q<1\).

In Fig.~\ref{fig:escapeex3}, we display the result of a system from
the third row of Fig.~\ref{fig:Rhat-grid}. It has $e_{1d}=0.2$,
$e_{2d}=0.1$, and an initial period ratio $P_2/P_1=1.5$.  At the
beginning of the integration, the two planets convergently migrate in
resonance.  Librations in the period ratio grow in amplitude over
time. Eventually, the planets escape the resonance. The inner planet
\(m_1\) is kicked outwards and \(m_2\) continues migrating inwards
until the two planets' orbits cross. This behavior could be related to
the instability discovered for an inner test particle in the MMR
capture \citep[][]{goldreich_overstable_2014,xu_migration_2018}.

\subsection{Test particle limit}
\label{sec:orga98868a}
\subsubsection{Relationship to the comparable mass case}
\label{sec:org26229a1}

The test particle Hamiltonian in equation~\eqref{eq:tpext} can also be
transformed into the form of equation~\eqref{hhat} through a canonical
shift, analagous to the reducing rotation utilized in Section
\ref{sec:org8116cb5}.  Relative to the comparable mass case, the
test-particle analysis is simpler because the vector \(\mathbf{e}_p\)
remains constant in time.  Because of this, we now return to the test
particle treatment of Section~\ref{sec:org4c72d92}, with the goal of
obtaining a heuristic understanding of apsidal alignment in real
planetary systems (i.e. \(0<q<\infty\)).

For simplicity, we restrict ourselves to the case of an exterior test
particle (as in Fig.~\ref{fig:tp-grid-ext}), which is equivalent to
the formal limit of \(q=m_1/m_2\to\infty\).  We now compare the
\(q=\infty\) case to the \(q=2\) case, the latter of which we
have investigated in Section~\ref{sec:orgd5c121f} (i.e. the
first two rows Fig.~\ref{fig:Rhat-grid}).  Traversing Figure
\ref{fig:tp-grid-ext} along the \(e_p\) axis is
equivalent to traversing the \(e_{1,\rm d}\) axis
in Fig.~\ref{fig:Rhat-grid}.  However, the \emph{vertical} axis of
Fig.~\ref{fig:tp-grid-ext} (\(h\)) is \emph{not} equivalent to the
vertical axis in Fig.~\ref{fig:Rhat-grid} (\(e_{2,\rm d}\)).  To
relate these two quantities, we use \(e_{\rm eq}\)
\(= h\sqrt{1.73/j}\) (equation~\ref{eq:eeqext}) as a proxy for the
``disk-driven eccentricity'' of the test particle.
One can see in
Fig.~\ref{fig:Rhat-grid} that, for the \(q=2\) systems that do not
escape, the division between ``\(\varpi-\text{aligned}\)'' and
``\(\varpi-\text{anti-aligned}\)'' roughly corresponds with the division
between ``\(\varpi-\text{aligned}\)'' and ``\(\varpi-\text{circulating}\)''
for the test particle systems in Fig.~\ref{fig:tp-grid-ext}. In the
following, we treat ``\(\varpi-\text{circulating}\)'' and
``\(\varpi-\text{anti-aligned}\)'' as being equivalent dynamical states.

\subsubsection{The condition for apsidal alignment}
\label{sec:org38a537a}

We again adopt the notation used in Section~\ref{sec:org4c72d92} for
\(q=\infty\).  The resonant angle \(\hat\theta\) can now be written as
\begin{align}
  \label{eq:tphhat}
  \tan\hat\theta = \frac{e\sin\theta}{e\cos\theta - \abs{f_1/f_2}e_p\cos\theta_p},
\end{align}
which is the test-particle limit of equation~\eqref{eq:hattheta} (see
Appendix~\ref{sec:orgf6e2109} for details),
where we have assumed \(\varpi_p=0\).  In both the
\(\varpi-\text{aligned}\) and $\varpi-\text{circulating}$ cases,
\(\theta_p\) circulates and \(\hat\theta\) (as written in equation
\ref{eq:tphhat}) librates around \(0^\circ\). However, \(\theta\)
librates around \(180^\circ\) when \(\varpi\) circulates, and it
circulates when \(\varpi\) librates around \(0\).  This can be
understood using the test-particle Hamiltonian (equation
\ref{eq:tpext}): \(\dot\varpi\) is proportional to \(\cos\theta\),
and so \(\varpi\) cannot librate if \(\theta\) remains close to
\(180^\circ\).

The angle \(\theta\) therefore determines the behavior
of \(\Delta\varpi\) in resonance for a test particle.
Consider equation~\eqref{eq:tphhat}: since \(\hat\theta\) always librates
around \(0\), then \(\tan\hat\theta\) remains finite;  this implies that
the denominator never reaches \(0\).  Because \(\theta_p\) always
circulates, \(\cos\theta_p\) takes on values between \(-1\) and \(1\).
Thus, when \(\theta\) librates around \(180^\circ\), the following
condition is required for \(\hat\theta\) to librate:
\begin{align}
  \label{eq:tpcond}
  -e + \left|\frac{f_1}{f_2}\right|e_p < 0.
\end{align}
We assume the eccentricity librates around a central eccentricity
\(e_c \simeq e_{\rm eq} + \Delta_e\), which defines the parameter
$\Delta_e$, and substitute this into equation~\eqref{eq:tpcond}:
\begin{align}
  e_{\rm eq} > \left|\frac{f_1}{f_2}\right|e_p - \Delta_e.
\end{align}
Using equation~\eqref{eq:eeqext} to relate \(h\) and \(e_{\rm eq}\),
we estimate the transition to aligned perihelia to occurs for
\(h<h_c(e_p)\), where \(h_c\) is given by
\begin{equation}
\label{eq:phasespaceline}
h_c(e_p) = \sqrt{\frac{j}{1.73}}\left(\left|\frac{f_1}{f_2}\right|e_p-\Delta_e\right)
\end{equation}
\noindent
In principle, $\Delta_e$ could be zero, but we find $\Delta_e=0.015$
to be a better approximation, which we use for the line in Figure
\ref{fig:tp-grid-ext}.  Despite the many approximations, this analysis
reproduces the numerical results fairly well.

\subsubsection{Discussion for \(0<q<\infty\)}
\label{sec:org13f48ed}
The above analysis shows that test particle
apsidal alignment can be understood as a competition between
$e_{\rm eq}$, which is related to $h$, and
\(e_p\), the eccentricity of
the massive planet.  Our interpretation of this result in the
comparable mass context is to identify the disc driven eccentricities \(e_{1,d}\) and \(e_{2,d}\)
of Section~\ref{sec:orgd5c121f} with \(e_{\rm eq}\) and \(e_p\) in Section~\ref{sec:org4c72d92}.

Apsidal behavior in the comparable mass regime is more complex than
the test-mass system.  We observe that \(|\Delta\varpi|=0\),
\(90^\circ\), and \(180^\circ\) are all possible.  Each planet's
eccentricity works on the other to anti-align their periapses in
resonance. However, if disc forces drive them \emph{away} from the
ratio \(e_1/e_2=1/q\), the tendency of anti-alignment can be overcome,
resulting in a different apsidal configuration.  The resulting
configuration has a complicated dependence on the planet and disk
parameters, with some systems escaping resonance (see
Fig.~\ref{fig:Rhat-grid}).  A detailed analysis of the effects of
different parameters \(q\), \(h\), \(T_e\), and \(e_{i,d}\) would
reveal transitions between the resonance modes analogous to equation
\eqref{eq:phasespaceline} for the comparable mass regime.

\section{Conclusion}
\label{sec:org71c2a7a}
We have studied the mean-motion resonance capture of two migrating
planets in protoplanetary disks, focusing on the property of the
apsidal angles of captured planets. Our study is motivated by recent
observations, which show that planets in MMR can be either apsidally
aligned or anti-aligned (see Section~\ref{sec:org493ee54}).

In the standard picture of MMR capture, planets undergo convergent
migration and experience eccentricity damping due to planet-disk
interactions.  We show in Section~\ref{sec:org985dec7} that this
standard picture always leads to capture where the resonance angles,
$\theta_1=(j+1)\lambda_2-j\lambda_1-\varpi_1$ and
$\theta_2=(j+1)\lambda_2-j\lambda_1-\varpi_2$, librate around zero or
$\pi$, and such capture produces apsidal anti-alignment
($\Delta\varpi=\varpi_1-\varpi_2=180^\circ$).

To explore the possibility of producing apsidal alignment in MMR
capture, we analyze the problem of a test particle in the vicinity of
an MMR with a planet of mass \(m_p\) (Section~\ref{sec:org4c72d92}).
We find that apsidal alignment occurs when the planet's eccentricity
$e_p$ is comparable or larger than the ``equilibrium'' eccentricity of
the test particle captured in MMR (see Fig.~\ref{fig:tp-grid-ext}),
the latter results from the migration and eccentricty damping by the
disc, and depends on the disc apsect ratio $h$ (see
Eq.~\ref{eq:dotedriving}).

Our test particle results inform our analysis of how apsidal alignment
may arise in the case of comparable-mass planets.  In
Section~\ref{sec:orgd5c121f}, we show that when the planets experience
eccentricity driving due to their interactions with the disc, apsidal
alignment in MMR capture can be produced. The eccentricity driving
forces prevent the libration of $\theta_1$ and $\theta_2$, allowing
the captured planets to settle into the apsidally aligned state (see
Fig.~\ref{fig:drivingex}) in which a ``mixed'' resonant angle
$\hat\theta$ librates.  However, in the presence of eccentricity
driving, the process of MMR capture is highly irregular; depending on
the initial condition, the planet mass ratio, and the magntitudes of
the driving forces, various outcomes can be produced, including
apsidal alignment, anti-alignment, and a perpendicular configuration
($\Delta\varpi=90^\circ$), as well as resonance disruption (see
Fig.~\ref{fig:Rhat-grid}).

This paper represents the first investigation into the effect of
eccentricity driving in mean-motion resonant systems.  The observed
apsidal alignment in the K2-19 system, the moderate eccentricites of
K2-19b and c, and the libration of the ``mixed'' resonant angle
\citep[$\hat\theta$;][]{petit_resonance_2020}, can all be produced by
this effect.  These suggest that the planets in the K2-19 system have
interacted with an eccentricity driving disc in the past.  In
addition, our finding that MMR capture can be disrupted by
eccentricity driving may also contribute to the observed
underabundance of exact MMRs (with $P_2/P_1 = (j+1)/j$) in the Kepler
multi-planet systems \citep{fabrycky_architecture_2014}, simply
because the exact resonant systems can be pushed to slightly larger
period ratios whenever the disc drives the planet eccentricities.


Our results come with the obvious caveat that we have used simple
parameterized models (with constant dissipative timescales) for the
planetary eccentricity damping and driving by the disc.  In reality,
the coupling between the disc and planet is a function of
eccentricity, location in the disk, and disc profile.
Long-term hydrodynamical simulations of two migrating planets in discs,
including the possibility of eccentricity driving, would 
be needed to fully explore the effects studied in this paper.

\section*{Acknowledgements}
This work is supported in part by NSF grant AST-2107796 and the
NASA grant 80NSSC19K0444.
\section*{Data Availability}
The code used to generate the data for this paper can be found on
https://github.com/jtlaune/mmr-apsidal-angle. All figures can be
reproduced from this data.

\bibliography{apsidal-alignment}
\bibliographystyle{mnras}
\clearpage
\onecolumn
\appendix
\section{Comparable mass Hamiltonian}
\label{sec:org2b22ae8}
\subsection{Scaling the Hamiltonian}
\label{sec:orgb93cb22}
The Hamiltonian for two comparable mass planets near the \(j:j+1\)
resonance is
\begin{align}
  H = -\frac{G M m_{1}}{2 a_{1}}-\frac{G M m_{2}}{2 a_{2}}
                 -\frac{G m_{1} m_{2}}{a_{2}}
                  \left[
                  f_{1} e_{1} \cos \theta_{1} 
                  +f_{2} e_{2} \cos \theta_{2}\right],
\end{align}
where $\theta_1=(j+1)\lambda_2-j\lambda_1-\varpi_1$ and
$\theta_2=(j+1)\lambda_2-j\lambda_1-\varpi_2$ (equations~\ref{circangles1} and
\ref{circangles2}) are the resonance angles and $f_1>0$, $f_2<0$
(equations~\ref{eq:coefficients1} and \ref{eq:coefficients2}) are
functions of the SMA ratio, $\alpha=a_1/a_2$.  Define
\(m_{\rm tot} = m_1+m_2\) and let \(a_0\) be the scale length of the
problem.  We will then scale the Hamiltonian by
\(H_0 = GMm_{\rm tot}/a_0\), the time by the frequency
\(\omega_0 = \sqrt{GM/a_0^3}\), and the canonical momenta by
\(\Lambda_0 = m_{\rm tot} \sqrt{GMa_0}\).  The dimensionless
Hamiltonian \(\mathcal{H}\) is then
\begin{align}
  \mathcal{H} \equiv \frac{H}{H_0}
  = -\frac{m_1/m_{\rm tot}}{2a_1/a_0}
    -\frac{m_2/m_{\rm tot}}{2a_2/a_0}
  -\frac{\tilde m}{M (a_2/a_0)}\left[
    f_1e_1\cos\theta_1+f_2e_2\cos\theta_2
    \right],
\end{align}

\noindent
where \(\tilde m = m_1m_2/m_{\rm tot}\) is the reduced mass. We assume
the reduced mass ratio is small ($\tilde\mu = \tilde m/M\ll1$).
The canonical momenta, coordinate pairs are
\begin{align}
  \Lambda_1 &= \frac{m_1}{m_{\rm tot}}\sqrt{\frac{a_1}{a_0}},\quad \lambda_1, \\
  \Lambda_2 &= \frac{m_2}{m_{\rm tot}}\sqrt{\frac{a_2}{a_0}},\quad \lambda_1, \\
  \Gamma_1 &= \frac{m_1}{m_{\rm tot}}\sqrt{\frac{a_1}{a_0}}
             \left(1-\sqrt{1-e_2^2}\right),\quad \gamma_1, \\
  \Gamma_2 &= \frac{m_2}{m_{\rm tot}}\sqrt{\frac{a_2}{a_0}}
             \left(1-\sqrt{1-e_2^2}\right),\quad \gamma_2,
\end{align}
where $\gamma_1=-\varpi_1$ and $\gamma_2=-\varpi_2$.  The Hamiltonian
can be expressed as a function of the momenta and resonance angles,
\begin{align}
\label{eq:H_1}
  \mathcal{H}
  = -\frac{q^3}{2(1+q)^3 \Lambda_1^2}
    - \frac{1}{2(1+q)^3\Lambda_2^2}
   - \frac{\tilde\mu}{(1+q)^2 \Lambda_2^2}\left[
    f_1\sqrt{\frac{2\Gamma_1}{\Lambda_1}}\cos\theta_1
    +f_2\sqrt{\frac{2\Gamma_2}{\Lambda_2}}\cos\theta_2
    \right],
\end{align}
where we have used $q=m_1/m_2$.

\subsection{Transforming the Hamiltonian}
\label{sec:org70f1898}
We would like to find the momenta conjugate to the fast coordinates
\(\lambda_i\) while making the slowly varying \(\theta_i\) conjugate
to $\Gamma_i$.  Such a canonical transformation preserves the form
\begin{align}
  \label{eq:dH} 
  d\mathcal{H}
  &= \Lambda_1 d\lambda_1+\Lambda_2d\lambda_2
    + \Gamma_1d\gamma_1+\Gamma_2d\gamma_2\nonumber\\
  &= \Gamma_1 d\theta_1 + \Gamma_2 d\theta_2
    +J_1 d\lambda_1+J_2d\lambda_2 .
\end{align}
We can solve the set of equations in \eqref{eq:dH} for
\begin{align}
\label{eq:J1}
J_1 &= \Lambda_1 + j(\Gamma_1+\Gamma_2),\\
\label{eq:J2}
J_2 &= \Lambda_2 - (j+1)(\Gamma_1+\Gamma_2),
\end{align}

\noindent where \(\Gamma_i\) and \(J_i\) are now conjugate to
\(\theta_i\) and \(\lambda_i\), respectively.
The coordinates \(\lambda_1\) and \(\lambda_2\)
no longer appear in the Hamiltonian,
which means \(J_1\) and \(J_2\) are constants of motion and
equation~\eqref{eq:H_1} may be written
in the following form:
\begin{align}
\label{eq:H_2}
  \mathcal{H}
  = \mathcal{H}_0(\Gamma_1+\Gamma_2; J_1, J_2, q)
                  + \mathcal{H}_{\rm pert}(\theta_1, \theta_2, \Gamma_1,\Gamma_2; J_1, J_2, q),
\end{align}

\noindent
where
\begin{align}
  \label{eq:H01}
  \mathcal{H}_0(\Gamma_1+\Gamma_2; J_1, J_2, q)
  = -\frac{q^3}{2(1+q)^3(J_1-j(\Gamma_1+\Gamma_2))^2}
  -\frac{1}{2(1+q)^3(J_2+(j+1)(\Gamma_1+\Gamma_2))^2} 
\end{align}

\noindent
and
\begin{align}
  \label{eq:Hpert1}
  \mathcal{H}_{\rm pert}(\Gamma_1,\Gamma_2; J_1, J_2, q)
  = -\frac{\tilde\mu}{(1+q)^2(J_2+(j+1)(\Gamma_1+\Gamma_2))^2}
  \left[
    f_1\sqrt{\frac{2\Gamma_1}{J_1 - j(\Gamma_1+\Gamma_2)}}\cos\theta_1
  +f_2\sqrt{\frac{2\Gamma_2}{J_2 + (j+1)(\Gamma_1+\Gamma_2)}}\cos\theta_2
    \right].
\end{align}

\noindent We have \(\Gamma_i \ll \Lambda_i\) for small
eccentricities.  Under this assumption, we may drop terms smaller than
\(\mathcal{O}(\Gamma_i^2/\Lambda_i^4)\).  Equation~\eqref{eq:H01} becomes
\begin{align}
  \label{eq:H02}
  \mathcal{H}_0
  = \mathcal C_0(q, J_1, J_2) -\frac{1}{(1+q)^3}\left[
     \frac{q^3}{2\Lambda_1^2} + \frac{1}{2\Lambda_2^2}
   + \left(
     \frac{jq^3}{\Lambda_1^3} - \frac{(j+1)}{\Lambda_2^3}
     \right)(\Gamma_1+\Gamma_2)
   -\frac32\left( 
     \frac{j^2q^3}{\Lambda_1^4} - \frac{(j+1)^2}{\Lambda_2^4}\right)
     (\Gamma_1+\Gamma_2)^2
     \right],
\end{align}

\noindent
where \(\mathcal C_0\) is a constant of resonance which depends on
initial conditions.  Hence, we leave \(\mathcal C_0\) out of the
following calculations.  Absent any dissipation, \(\Lambda_1\) and
\(\Lambda_2\) are approximately constant in resonance.  Hence, we may
also drop the first two terms in parentheses in
equation~\eqref{eq:H02}, leaving only the terms which include factors
of \((\Gamma_1+\Gamma_2)\):
\begin{align}
  \label{eq:H03}
  \mathcal{H}_0
  = -\frac{1}{(1+q)^3}\left[
   \left(
     \frac{jq^3}{\Lambda_1^3} - \frac{(j+1)}{\Lambda_2^3}
     \right)(\Gamma_1+\Gamma_2)
   -\frac32\left( 
     \frac{j^2q^3}{\Lambda_1^4} - \frac{(j+1)^2}{\Lambda_2^4}\right)
     (\Gamma_1+\Gamma_2)^2
     \right].
\end{align}
The perturbation part (equation~\ref{eq:Hpert1})
reduces to its original form,
\begin{align}
\label{eq:H_3}
  \mathcal{H}_{\rm pert}
  =-\frac{\tilde\mu}{(1+q)^2\Lambda_2^2}
  \left[
  f_1\sqrt{\frac{2\Gamma_1}{\Lambda_1}}\cos\theta_1
  +f_2\sqrt{\frac{2\Gamma_2}{\Lambda_2}}\cos\theta_2
  \right],
\end{align}
because there is already a small factor ($\tilde\mu$) in the
numerator.

\subsection{Reducing rotation}
Following \citet{henrard86_reduc_trans_apocen_librat}
\citep[equivalently,
][]{wisdom_canonical_1986,deck13_first_order_reson_overl_stabil,moutamid14_coupl_between_corot_lindb_reson},
let \(\v X\) be the Cartesian formulation
\begin{align}
  \v X &= (x_1, x_2, X_1, X_2)\nonumber\\
  &= (\sqrt{\Gamma_1}\cos\theta_1, \sqrt{\Gamma_2}\cos\theta_2,
    \sqrt{\Gamma_1}\sin\theta_1, \sqrt{\Gamma_2}\sin\theta_2)
\end{align}

\noindent 
Define
\begin{align}
    g_1 &= f_1\sqrt{\frac{2}{\Lambda_1}}, \\
    g_2 &= f_2\sqrt{\frac{2}{\Lambda_2}}, \\
\end{align}

\noindent and
\begin{align}
  \mathcal{A} = \frac{1}{\sqrt{g_1^2+g_2^2}}.
\end{align}

\noindent The perturbation Hamiltonian, \(\mathcal H_{\rm
pert}\) (equation~\ref{eq:H_3}), has
\begin{align}
  \mathcal H_{\rm pert} \propto g_1 x_1 + g_2 x_2.
\end{align}
Let \(\v \Psi\) be the counter-clockwise phase space rotation by the
angle \(\psi\), where \(\tan\psi= g_2/g_1\),
\begin{align}
  \v \Psi =  \mathcal{A}
  \begin{pmatrix}
    g_1 & g_2 \\
    -g_2 & g_1 
  \end{pmatrix}.
\end{align}

\noindent The block matrix
\begin{align}
  \v M =
  \begin{pmatrix}
    \v \Psi & \v 0 \\
    \v 0 & \v \Psi
  \end{pmatrix}
\end{align}

\noindent is symplectic \citep{goldstein_classical_2000}.
The coefficients \(g_i\) depend weakly on the semimajor axis ratio
\(\alpha\), and so \(\v M\) only represents a canonical transformation if
\(\alpha\) is stationary or varying slowly, which is a good
approximation for the systems considered in this paper.

Define the coordinates
\begin{align}
   \v W = (w_1, w_2, W_1, W_2) \equiv \v M \v X.
\end{align}

\noindent so that \(w_1 = \mathcal{A}(g_1 x_1 + g_2 x_2)\).  Hence,
\(\mathcal H_{\rm pert}\propto w_1\) only.  Finally, we revert the
\(\v W\) set back to polar coordinates
\((\hat\theta,\hat\theta_2,S_1',S_2')\), so that
\(\mathcal H_{\rm pert}\propto \sqrt{S'_1}\cos\hat\theta\) only.  The sum
\begin{align}
  \Gamma_1 +\Gamma_2 = x_1^2+x_2^2 + X_1^2 + X_2^2
  = w_1^2+w_2^2 + W_1^2 + W_2^2 = S'_1 + S'_2
\end{align}
is preserved, and so the form of \(\mathcal H_0\)
is preserved:
\begin{align}
  \label{eq:H03}
  \mathcal{H}_0
  = -\frac{1}{(1+q)^3}\left[
   \left(
     \frac{jq^3}{\Lambda_1^3} - \frac{(j+1)}{\Lambda_2^3}
     \right)(S'_1 + S'_2)
   -\frac32\left( 
     \frac{j^2q^3}{\Lambda_1^4} - \frac{(j+1)^2}{\Lambda_2^4}\right)
     (S'_1+S'_2)^2
     \right].
\end{align}
The perturbation part is now
\begin{align}
  \mathcal{H}_{\rm pert}
  =-\frac{\tilde\mu}{(1+q)^2\Lambda_2^2} \sqrt{S'_2}\cos\hat\theta.
\end{align}
The new resonance angle is
given by the equation
\begin{align}
\label{hattheta}
  \tan\hat{\theta} = \frac{W_1}{w_1} = \frac{f_1 e_1\sin(\theta_1)
  + f_2e_2\sin(\theta_2)}{f_1e_1\cos(\theta_1) + f_2e_2\cos(\theta_2)}
\end{align}
and is conjugate to the momentum
\begin{align}
  S_1' = w_1^2 + W_1^2 = \mathcal{A}^2(f_1^2e_1^2
  + 2f_1f_2e_1e_2\cos(\varpi_1 - \varpi_2) + f_2^2e_2^2).
\end{align}
After the reducing rotation, neither \(\mathcal H_0\) nor \(\mathcal
H_{\rm pert}\) depend on \(\hat\theta_2\), and so its conjugate momentum,
\(S_2\), is a constant of resonance:
\begin{align}
\label{eq:appS2deriv}
  S_2' = w_2^2 + W_2^2 = \mathcal{A}^2\left(q\sqrt{\alpha}f_2^2e_1^2
-2f_1f_2e_1e_2\cos(\varpi_1-\varpi_2) + \frac{f_1^2}{q\sqrt\alpha}e_2^2\right).
\end{align}

Through a scale transformation (which changes the time units by a
factor of $1/\mathcal A^2$), we can remove the $\mathcal{A}^2$ term in
$S'_1$ and $S'_2$. This results in the momentum
\begin{align} \hat R = \frac{1}{\mathcal{A}^2} S'_1 = f_1^2e_1^2
  - 2|f_1f_2|e_1e_2\cos(\varpi_1 - \varpi_2) + f_2^2e_2^2
\end{align}
and conserved quantity
\begin{align}
  S_2 = \frac{1}{\mathcal{A}^2} S'_2 = q\sqrt{\alpha}f_2^2e_1^2
+2|f_1f_2|e_1e_2\cos(\varpi_1-\varpi_2) + \frac{f_1^2}{q\sqrt\alpha}e_2^2,
\end{align}
where we have used the fact $f_1f_2=-|f_1f_2|$.  \(\hat R\) has the
following geometric interpretation:
\begin{align}
  \hat R = \left\lvert \abs{f_1}\mathbf{e}_1 - \abs{f_2}\mathbf{e}_2\right\rvert^2,
\end{align}

\noindent
where the \(\mathbf{e}_i\) are the Runge-Lenz vectors with magnitude
\(e_i\) in the direction of \(\varpi_i\).  The key characteristic of
the $\hat R$, $\hat \theta$ conjugate pair is that it does not depend
on the planetary mass ratio
\citep[e.g.][]{deck13_first_order_reson_overl_stabil}.

Altogether, we arrive at the following Hamiltonian after
dropping constant terms:
\begin{align}
  \label{eq:HShat}\mathcal H(\hat \theta, \hat R) &= \mathcal H_0(\hat R) + \mathcal H_{\rm pert}(\hat \theta, \hat R), \\
  \mathcal H_0
  &= \left( 3\mathcal N\mathcal{A}^2S_2 -\mathcal M\right) \mathcal A^2 \hat R
    - \frac32 \mathcal N \mathcal A^4\hat R^2, \\
  \mathcal H_{\rm pert}
  &= - \tilde\mu\mathcal K\mathcal A\sqrt{\hat R}\cos\hat\theta,
\end{align}
with coefficients given by
\begin{align}
  \mathcal M
  &= \frac{1}{(1+q)^3}\left(\frac{jq^3}{\Lambda_1^3}-\frac{j+1}{\Lambda_2^3}\right),\\
  \mathcal N
  &= \frac{1}{(1+q)^3}\left(\frac{j^2q^3}{\Lambda_1^4}-\frac{(j+1)^2}{\Lambda_2^4}\right),\\
  \mathcal K
  &= \frac{1}{(1+q)^2\Lambda_2^2}.
\end{align}

\section{Test particle Hamiltonian}
\label{sec:orgf6e2109}
The internal and external test particle limits are largely analagous,
and so we focus on the external limit, with \(q\) approaching
infinity.  The inner planet now has constant \(m_1=m_p>0\),
\(a_1=a_p\), \(e_1=e_p\), and \(\varpi_1=\varpi_p\). We may
arbitrarily set \(\varpi_p=0\) due to rotational symmetry.  The test
particle has \(m_2=0\), \(a_2=a\), \(e_2=e\), and \(\varpi_2=\varpi\).
We are again neglecting dissipative and secular effects in our
analysis.  After transforming to the dimensionless Poincair\'e
elements \(\Lambda=\sqrt{a/a_p}\) and
\(\Gamma=\Lambda(1-\sqrt{1-e^2})\approx \frac12\Lambda e^2\), the
Hamiltonian is
\begin{align}
\label{eq:tpresH}
  \mathcal{H}
  &= - \frac{1}{2\Lambda^2} - \mu_p\left(f_2
    \sqrt{\frac{2\Gamma}{\Lambda}}\cos(\theta_p+\gamma) + f_1 e_p\cos\theta_p\right),
\end{align}

\noindent
where \(\theta_p = (j+1)\lambda - jn_pt\) is now an explicit
function of time. The coordinate \(\gamma=-\varpi\) is conjugate
to \(\Gamma\). Utilizing the approximation that \(\alpha\) is varying
adiabatically, we can effectively treat \(\Lambda\) and \(f_i\) as
constants while in resonance. 

Because we are now dealing with a potential that is an explicit
function of time, the test particle formulation is formally different
than the comparable mass problem, which is defined only for the
parameter range \(0<q<\infty\).  The canonical transformation described
in Appendix \ref{sec:org2b22ae8}, which is generated by the infinitesimal rotation
about the origin, must also formally change.  The corresponding
transformation here is generated by the infinitesimal translation
towards the inner planet's Runge-Lenz vector, \(\mathbf{e_p}\).

We will reduce equation~\eqref{eq:tpresH} to a single
degree of freedom Hamiltonian. If \(e_p=0\), the resonance
angle is \(\theta=\theta_p+\gamma\). For \(e_p>0\), we will
derive an angle \(\hat\theta\), analagous to 
equation~\eqref{hattheta}, which incorporates the value \(e_p\).
Similar to our comparable mass derivation,
we first switch to Cartesian coordinates
\(\mathbf X = (X,Y) = (\sqrt{\Gamma}\cos\gamma, \sqrt{\Gamma}\sin\gamma)\)
so that the Hamiltonian now becomes
\begin{align} 
  \mathcal{H}
  &= - \frac{1}{2\Lambda^2} - \mu_p\left(f_2
    \sqrt{\frac{2}{\Lambda}}X\cos\theta_p + f_2\sqrt{\frac{2}{\Lambda}}Y\sin\theta_p + f_1 e_p\cos\theta_p\right).
\end{align}
The first and third term now have identical dependence on
\(\theta_p\).  The transformation of
\(\mathbf X\) coordinates given by
\begin{align}
  X' &= X + \frac{f_1e_p}{f_2}\sqrt{\frac{2}{\Lambda}}, \\
  Y' &= Y,
\end{align}

\noindent
induces a new canonically conjugate pair \(\mathbf X'=(X',Y')\).
The Hamiltonian becomes
\begin{align}
  \mathcal{H}
  &= - \frac{1}{2\Lambda^2} - \mu_p\left(f_2
    \sqrt{\frac{2}{\Lambda}}X'\cos\theta_p + f_2\sqrt{\frac{2}{\Lambda}}Y'\sin\theta_p \right).
\end{align}

\noindent
Finally, returning back to the canonical polar coordinates,
\begin{align}
  \tan\hat\gamma
  &= \frac{Y'}{X'} \\
  &= \frac{f_2e\sin\gamma}{f_2e\cos\gamma + f_1e_p},\label{eq:hatgtp}\\
  \hat\Gamma
  &= X'^2 + Y'^2 \\
  &= \Gamma + \frac{f_1e_p}{f_2}\sqrt\frac{\Lambda\Gamma}{2}
    + \frac{f_1^2}{f_2^2}\frac{\Lambda e_p^2}{2},\label{eq:hatGtp}\\
\end{align}

\noindent
we may write the Hamiltonian as
\begin{align}
  \mathcal{H}
  &= - \frac{1}{2\Lambda^2} - \mu_p\left(f_2
    \sqrt{\frac{2\hat\Gamma}{\Lambda}}\cos(\theta_p+\hat\gamma) \right).
\end{align}

\noindent
The derivation may now continue as if this were the CR3BP,
which culminates with the following action-angle pair:
\begin{align}
\hat R =
\frac{2f_2^2}{\Lambda}\hat\Gamma
  &=f_1^2 e_p^2 + 2 f_2 f_1 e_pe\cos\hat\gamma + f_2^2 e^2,  \\
\hat\theta
  &= \theta_p + \hat\gamma.
\end{align}

\end{document}